\documentclass[11pt,a4,epsfig]{article}
\usepackage{amssymb}   

\begin{titlepage}
\end{titlepage}
\normalsize
\usepackage{graphicx}

\def\mb#1{\setbox0=\hbox{$#1$}\kern-.025em\copy0\kern-\wd0
\kern-0.05em\copy0\kern-\wd0\kern-.025em\raise.0233em\box0}

\begin{document}

\title{Trapping of Dust by Coherent Vortices in the Solar Nebula}
\author{ Pierre-Henri CHAVANIS{$^{1,2}$} \\
\\
{\small 
\parbox{11.5cm}{$^1$\parbox[t]{11cm}{ Laboratoire de Physique Quantique,
Universit\'e Paul Sabatier, 118 route de Narbonne 31062 Toulouse, France }
$^2$\parbox[t]{11cm}{Istituto di Cosmogeofisica, Corso Fiume 4, 10133 Torino,
Italia}
}}}
\date{\today}
\maketitle

\begin{abstract}

We develop the idea proposed by Barge \& Sommeria (1995) and Tanga {\it et al.}
(1996) that large-scale vortices present in the solar nebula can concentrate
dust particles and facilitate the formation of planetesimals and planets. We
introduce an exact vortex solution of the incompressible 2D Euler equation and
study the motion of dust particles in that vortex. In particular, we  derive
analytical expressions for the capture time and the mass capture rate as a
function of the friction parameter. Then, we study how small-scale turbulent
fluctuations affect the motion of the particles in the vortex and determine
their rate of escape by solving a problem of quantum mechanics. We apply these
results to the solar nebula and find that the capture is optimum near Jupiter's
orbit (as noticed already by Barge \& Sommeria 1995) but also in the Earth
region. This second optimum corresponds to the transition between the Epstein
and the Stokes regime which takes place, for relevant particles, at the
separation between telluric and giant planets (i.e near the asteroid belt). At
these locations, the particles are efficiently captured and concentrated by the
vortices and can undergo gravitational collapse to form the planetesimals.

   \end{abstract}

\section{ Introduction}
\label{sec_intro}

Many astrophysical objects, ranging from young stars to massive black holes, are
surrounded by widespread gaseous disks. The existence of a primordial disk
around the sun was conjectured by Kant (1755) and Laplace (1796) in the
$18^{th}$ century to explain the quasi-circular, coplanar and prograd motion of
the planets in the solar system. Such protoplanetary disks have recently been
observed with the Hubble Space Telescope in the Orion nebula around stars less
than one million years old. These gaseous disks can be considered as a
by-product of the star formation: after the gravitational collapse of a
molecular cloud, about $99\%$ of the initial angular momentum is spread in an
extended disk, while $99\%$ of the mass forms the star, whose internal structure
is hardly affected by rotation.

Whenever it has been possible to observe rotating, turbulent fluids with good
resolution, it has been seen that individual, intense vortices form (Bengston \&
Lighthill 1982; Hopfinger {\it et al.} 1983; Dowling and Spiegel 1990). One of
the most striking example is Jupiter's Great Red Spot, a huge vortex persisting
for more than three centuries in the upper atmosphere of the planet. These
coherent vortices are well reproduced in numerical simulations (McWilliams 1990,
Marcus 1990) and laboratory experiments (Van Heist \& Flor 1989, Sommeria,
Meyers \& Swinney 1988) of two-dimensional turbulence and their organization can
be explained in terms of statistical mechanics (Miller 1990, Robert \& Sommeria
1991, Sommeria {\it et al.} 1991, Chavanis \& Sommeria 1998) \footnote {It can
be noted that the statistical mechanics of two-dimensional turbulence is very
similar to the theory of ``violent relaxation'' (Lynden-Bell 1967) by which
galaxies achieve an equilibrium state. It is fascinating to realize that despite
their very different physical nature, two-dimensional vortices and stellar
systems share some common features. This analogy is developed in detail in
Chavanis {\it et al.} (1996) and Chavanis (1996,1998a,1999).}.  It seems
therefore natural to expect their presence in accretion disks also (Dowling \&
Spiegel 1990, \ Abramowicz {\it et al.} 1992, Adams \& Watkins 1995). 

However, accretion disks are exceptional among rotating turbulent objects in the
strong shears that these bodies are believed to possess and this shear might
lead to rapid destruction of any structures that tend to form.  This objection
has been overruled by the numerical simulations of a two-dimensional flow in an
external Keplerian shear by Bracco {\it et al.} (1998,1999) for an
incompressible flow and Godon \& Livio (1999a,b,c) for a compressible flow.
Although cyclonic fluctuations are rapidely elongated and destroyed by the
shear, anticyclonic vortices form and persist for a long time before being
ultimately dissipated by viscosity. Naturally, this does not prove that coherent
structures must form on disks, but this strengthens the argument that disks are
likely to follow the norm of rotating, turbulent bodies. Other numerical results
(Hunter \& Horak 1983) and experimental work (Nezlin \& Snezhkin 1993) comfort
this point.

Coherent vortices in circumstellar disks can play an important role in the
transport of dust particles and in the process of planet formation.  Planets are
thought to be formed from the dust grains embedded in the disk after a
three-stage process: (i) in a first stage, microscopic particles suspended in
the gas stick together on contact due to electrostatic or surface forces. When
they reach sufficient sizes, they begin to sediment in the mid-plane of the disk
due to the combined effect of the gravity and the friction with the gas. When
settling dominates, a particle can grow by sweeping up smaller ones (Safronov
1969) and may easily reach sizes of several centimeters in a few thousand
orbital periods (Weidenschilling \& Cuzzi 1993). Bigger aggregates ($>100 cm$)
are more difficult to form on relevant time scales because of collisional
fragmentation (ii) When the layer of sedimented particles is sufficiently dense,
the gravitational instability is triggered and the layer crumbles into numerous
kilometer-sized bodies, the so-called ``planetesimals'' (Safronov 1969,
Goldreich \& Ward 1973). (iii) The subsequent evolution is marked by planet's
growth due to the accumulation of planetesimals in successive collisions. This
stage is well reproduced by dynamical models (Safronov 1969, Barge \& Pellat
1991, 1993). When the solid core becomes sufficiently massive, it can accrete
the surrounding gas and a giant planet, like Jupiter, is formed.

However, the above scenario faces two major problems. Recent studies have shown
that circumstellar disks are relatively turbulent and that small-scale
turbulence strongly reduces the sedimentation of the dust particles in the
ecliptic plane (Weidenschilling 1980, Cuzzi {\it et al.} 1993, Dubrulle {\it et
al.} 1995). For particles of relevant size, the density of the dust layer is not
sufficient to overcome the threshold imposed by Jeans instability criterion.
Therefore, the formation of the planetesimals, i.e the passage from $cm$-sized
to $km$-sized particles, is not clearly understood.  In addition, it seems
difficult, with the above model, to build up sufficiently massive cores in less
than one million years before the gas has been swept away by the solar wind
during the T-tauri phase (Safronov 1969, Wetherill 1988).  

Both difficulties are ruled out if we allow for the presence of vortices in the
disk. Their existence was first proposed by Von Weiz\"acker (1944) to explain
the regularity of the planet distribution in the solar system: the famous
Titius-Bode law \footnote{An account of Von Weizs\"acker's theory can be found
in Chandrasekhar (1946). Note that the idea of vortices in the solar system goes
back to Descartes (1643)}. This idea has been reintroduced recently by Barge \&
Sommeria (1995) and Tanga {\it et al.} (1996) who demonstrated that anticyclonic
vortices in a rotating disk are able to capture and concentrate dust particles.
The capture is made possible by  the action of the Coriolis force which pushes
the particles inward. These results are supported by a dynamical model which
integrates the motion of the particles in the velocity field produced by a full
Navier-Stokes simulation of the gas component (Bracco {\it et al.} 1999, Godon
\& Livio 1999c). It is found that the particles are very efficiently captured
and concentrated by the vortices. This is interesting because, without a
confining mechanism, $cm$-sized bodies would rapidely fall onto the sun due to
the inward drift associated with the velocity difference between gas and
particles. Inside the vortices, the density of the dust cloud is increased by a
large factor which is sufficient to trigger {\it locally} the gravitational
instability and facilitate the formation of the planetesimals or the cores of
giant planets. This trapping mechanism is quite rapid (a few rotations) and can
reduce substantially the time scale of planet formation.

In this paper, we present a simple analytical model for the capture of dust
particles by coherent vortices in a rotating disk. This model is directly
inspired by the numerical studies of Barge \& Sommeria (1995) and Tanga {\it et
al.} (1996) and their main results are recovered and confirmed. One interest of
our approach is to provide analytical results (leading to quantitative
predictions) and to isolate relevant parameters which prove to be particulary
important in the problem. In section \ref{sec_deterministic}, we introduce an
exact solution of the incompressible 2D Euler equation appropriate to our sudy.
This is an elliptic vortex with uniform vorticity matching continuously with the
azimuthal Keplerian flow at large distances. We consider deterministic
trajectories of dust particles in that vortex and derive analytical expressions
for the capture time and the mass capture rate as a function of the friction
parameter. In section \ref{sec_stochastic}, we investigate the effect of
small-scale turbulence on the motion of the particles. Their trajectories become
stochastic and their motion must be described in terms of diffusion equations.
We estimate the diffusion coefficient and determine the typical length on which
the particles are concentrated in the vortices. In section \ref{sec_escape}, we
evaluate the rate of particles which diffuse away from the vortices due to
turbulent fluctuations. An explicit expression for the ``rate of escape'' is
obtained by solving a problem of quantum mechanics, namely a two-dimensional
oscillator in a box. In appendix A, we give some details about the construction
of the vortex solution and in appendix B, we extend Toomre instability criterion
(Toomre, 1964) to the case of a {\it turbulent} rotating disk.

In parallel, we apply these theoretical results to the solar nebula and make
speculations about its actual structure. For relevant particles going from $10$
cm to $100$ cm in size, we remark that the transition between the Stokes and the
Epstein regimes (at which the gas drag law changes) corresponds precisely to the
transition between telluric (inner) and giant (outer) planets. Moreover, in each
zone there is a prefered location where the capture of dust by vortices is {\it
optimal}. For particles of density $\rho_{s}\sim 2$ g$/$cm$^3$ and size $\sim
30$ cm  (a typical prediction of grain growth models), this is near the Earth
orbit in the Stokes (inner) zone and between Jupiter and Saturn in the Epstein
(outer) zone. For a broader class of parameters, the prefered locations cover
the whole region of telluric and giant planets with a depletion near the
asteroid belt. Inside the vortices, the surface density of the dust particles is
increased by a factor $100$ or more sufficient to trigger {\it locally} the
gravitational instability. More precisely, we study how the surface density
enhancement depends on the size of the particles. We show that particles must
have reached at least centimetric sizes to collapse and form the planetesimals.
Smaller particles diffuse away from the vortices on account of turbulent
fluctuations  and are not sufficiently concentrated. This implies that sticking
processes are necessary to produce large particles. These results rehabilitate
the Safronov-Goldreich-Ward scenario in localized regions of the disk (i.e,
inside the vortices) and for sufficiently large (decimetric) particles .
Preliminary results of this work were presented in Chavanis (1998b).

\section{Deterministic motion of a particle in a vortex}
\label{sec_deterministic}

\subsection{The solar nebula model}
\label{sec_nebulamodel}

We assume that the solar nebula is disk-shaped and is in hydrostatic equilibrium
in the vertical direction. If the mass of the disk is much less that the solar
mass ($<0.1 M$), then its self-gravity can be neglected compared with the sun's
attraction. In that case, the disk has approximately Keplerian rotation
\begin{equation}
\Omega (r)=\biggl ({GM\over r^{3}}\biggr )^{1/2}
\label{OmegaKepler}
\end{equation} 
where $r$ is the distance from the sun and $G=6.672\ 10^{-8}$
cm$^{3}$/(g.s$^{2}$) the constant of gravity. For thin disks, the vertical
component of the solar gravity is:
\begin{equation}
g_{z}\simeq -{GM\over r^{2}} \times {z\over r}=-\Omega^{2}z
\label{gz}
\end{equation}  
The condition of hydrostatic equilibrium implies that the vertical pressure
gradient is precisely balanced by $g_{z}$, i.e:  
\begin{equation}
{\partial p\over\partial z}=-\rho_{g}\Omega^{2}z
\label{balancenebula}
\end{equation} 
If the local temperature $T={m\over k}{p\over\rho_{g}}$ is independant of $z$,
then the vertical density profile of the gas is 
\begin{equation}
\rho_{g}=\rho_{0}e^{-z^{2}/H^{2}}
\label{rho}
\end{equation} 
The half-thickness (or scale height) of the disk is given by
\begin{equation}
H=c_{s}/\Omega
\label{Hnebula}
\end{equation}
where $c_{s}\sim \sqrt{kT\over m}$ is the sound speed. Other plausible
temperature profiles, e.g, adiabatic gradient in the $z$ direction yield similar
results.

We assume also that the gas is turbulent. Turbulence is necessary to explain the
dynamical evolution of the protoplanetary disk and its cooling consistent with
cosmochemical data (Dubrulle 1993). However, its origin is not well understood
and remains controversial. Various mechanisms for inducing global nebula
turbulence have been proposed.  Lin \& Papaloizou (1980) and Cabot {\it et al.}
(1987a,b) suggested that thermal convection drives nebula turbulence. However,
the applicability of their results depends on the presence of abundant
micron-sized dust to provide substantial thermal opacity; thus, they are
questionable when significant particle growth has already occured and the
thermal opacity has decreased. Dubrulle (1992,1993) suggested that the Keplerian
shear is the main engine of the turbulence. This was already pointed out by Von
Weizs\"acker and further discussed by Chandrasekhar (1949) in view of the very
small viscosity of the disk: ``The successive rings of gas in the medium will
have motions relative to one another, and turbulence will ensue''. This
apparently obvious claim is actually far from straightforward to support because
the  Keplerian shear is stable with respect to linear, infinitesimal,
perturbations which are usually relied upon to induce turbulence. However,
Dubrulle \& Knobloch (1992) point out that it might be unstable to nonlinear
finite amplitude perturbations, a property shared by most of the shear flows
commonly met in engineering or laboratory experiments. The simplest example is
the plane Couette flow, a plane parallel stream of constant shear. This analogy
is contested by Balbus {\it et al.} (1996)  who didn't evidence nonlinear
instabilities in Keplerian disks at the respectable Reynolds numbers they
achieved numerically \footnote{See also the 2D simulations by Godon \& Livio
(1999a) starting from finite perturbations who showed that turbulence is not
self-sustained.}. These authors have suggested, in contrast, that turbulence in
Keplerian disks could be produced by a powerful MHD instability (Balbus \&
Hawley 1991). The numerical simulations of Bracco {\it et al.} (1998) suggest
that MHD turbulence can form magnetized vortices able to capture charged
particles. However, in the case of the disk that is supposed to have spawned the
solar system, it is thought that the matter was too cool to be ionized. The
recourse to magnetic effects to render the disk turbulent is therefore suspect
and the problem of whether Keplerian disks are turbulent or not remains open.
Recently, Lovelace {\it et al.} (1999) discoved a linear nonaxisymmetric
instability in a thin Keplerian disk which can lead to the formation of Rossby
vortices. This may open new perspectives for hydrodynamical turbulence in
accretion disks. 

In any case, the solar nebula must have been turbulent at least during its
formation from the collapsing protostar, because of velocity discontinuities as
the infalling matter struck the disk. The infall probably did not stop suddenly
but decayed over some interval. Therefore, the initial conditions in the disk
were turbulent and this is sufficient to form vortices that survive for many
rotation periods of the disk (Bracco {\it et al.} 1998,1999; Godon \& Livio
1999a,b,c). The appearance of large-scale coherent vortices in rotating flows is
due to the presence of the Coriolis force and the bimodal nature of turbulence
(Dubrulle \& Valdettaro 1992). At small scales, the influence of the rotation is
negligible and the turbulence is homogeneous and isotropic. Energy cascades
towards smaller and smaller scales up to the dissipation length. Turbulent
diffusion is important and can accelerate the mixing of dust particles. At
larger scales, the Coriolis force becomes dominant and the gas dynamics is quasi
two-dimensional. On these scales, the energy transfers are inhibited or even
reversed;  this is a manifestation of the celebrated ``inverse cascade'' process
(see, e.g, Kraichnan \& Montgomery 1980) in two-dimensional turbulence. There is
therefore the possibility of formation and maintenance of coherent vortices (Mc
Williams 1990). These vortices can form either by a large-scale instability
(Frisch {\it et al.} 1987, Dubrulle \& Frisch 1991, Kitchatinov {\it et al.}
1994) or by the succesive mergings of like-sign vortices, as observed in the
simulations of Bracco {\it et al.} (1998,1999) and Godon \& Livio (1999a,b,c).
Due to the strong Keplerian shear, only anticyclonic vortices can survive this
process.  Initially, the vortices have size $R\ll H$ and typical vorticity
$\Omega$. Their velocity $v\sim\Omega R$ is less that the sound speed
$c_{s}=\Omega H$ and the flow can be considered as incompressible. The merging
ends up when the Mach number $M_{a}={v\over c_{s}}$ reaches unity, i.e $R\sim
H$. Bigger vortices radiate density waves and do not maintain (see Barge \&
Sommeria 1995). 

We expect therefore that, after some evolution, the disk be filled with
anticyclonic vortices of typical size $H$, the disk thickness. Vortices are
expected to form throughout the nebula and there is no reason, a priori, to
beleive that certain regions of the disk should be excluded. However, two
vortices at comparable distance from the sun will approach each other (due to
differential rotation) and finally merge. Therefore, we do not expect more than
one (or a few) vortices at each radial distance. On the other hand, two
successive vortices should be separated by a distance comparable to their size
$R$. Since $R\sim H$ and $H$ scales like a power law of the distance to the sun
($r^{5/4}$ in the standard model considered below), the distribution of vortices
should be consistant with an approximate geometric progression of the planetary
positions (Barge \& Sommeria 1995). As elucidated by Graner \& Dubrulle (1994),
the Titius-Bode law reflects more the general properties of scale invariance in
the solar nebula than any particular physical process.

\subsection{The vortex  model}
\label{sec_vortexmodel}

We consider the motion of a dust particle in a vortex located at a distance
$r_{0}$ from the sun. For convenience, we work in a frame of reference rotating
with constant angular velocity $\Omega\equiv \Omega(r_{0})$ and we denote by
${\bf u}({\bf r},t)$ the velocity field of the gas in that frame.  The dust
particle is subjected to the attraction of the sun $-{GM\over r^{3}}{\bf r}$ and
to a friction with the gas that we write $-\xi ({\bf v}-{\bf u}({\bf r},t))$
where ${\bf v}={d{\bf r}\over dt}$ is the particle velocity. We shall come back
to this expression and to the value of the friction coefficient $\xi$ in section
\ref{sec_application1}. Since we work in a rotating frame, apparent forces arise
in the system. These are the Coriolis force $-2{\bf\Omega}\wedge {\bf v}$ and
the centrifugal force $\Omega^{2}{\bf r}$. All things considered, the particle
equation writes:
\begin{equation}
{d^{2}{\bf r}\over dt^{2}}=-\xi \biggl ({d{\bf r}\over dt}-{\bf u}({\bf
r},t)\biggr )-2{\bf \Omega}\wedge {d{\bf r}\over dt}+\biggl (\Omega^{2}-{GM\over
r^{3}}\biggr ){\bf r}
\label{motiongeneral}
\end{equation} 
This is an ordinary second order differential equation coupled to the gas
motion. The case of a time dependant velocity field ${\bf u}({\bf r},t)$
produced by the Navier-Stokes simulation of a random initial vorticity field
superposed on the Keplerian rotation has been investigated by Bracco {\it et
al.} (1999) and Godon \& Livio (1999c) who observed the capture of the dust
particles by anticyclonic vortices. The case of a static velocity field ${\bf
u}({\bf r})$ was first considered by Barge \& Sommeria (1995) and Tanga {\it et
al.} (1996) with different vortex profiles. Barge \& Sommeria (1995) assume that
the velocity field inside the vortex is made of concentric epicycles while it
matches continuously the Keplerian flow at large distances. This epicyclic model
is motivated by the underlying idea that the particles, when sufficiently
concentrated, will retroact on the vortex and will force the gas to follow their
natural motion. Tanga {\it et al.} (1996) consider a more complex streamfunction
describing an ensemble of small vortices corresponding to Rossby waves
corotating with the flow. This velocity field is obtained as a self-similar
solution of the linearized equations of motion governing the large-scale
dynamics of a turbulent nebula. These two models lead to qualitatively similar
results indicating that dust particles are efficiently trapped into coherent
anticyclonic vortices. However, the models differ in the details: in Tanga {\it
et al.} (1996), the particles sink to the center of the vortices with no limit
while in Barge \& Sommeria (1995), the spiralling motion ends up on an epicycle.
In that case, the friction force cancels out and the epicyclic motion is an
exact solution of the particle equation (\ref{motiongeneral}).

We must note, however, that the vortex constructed by Barge \& Sommeria (1995)
is relatively  {\it ad hoc} and does not satisfy the fluid equations rigorously.
In this  article, we introduce an exact vortex solution of the incompressible 2D
Euler equation and study analytically the motion of dust particles in that
vortex. Since the vortices of the solar nebula are small compared with the
radial distance $r_{0}$ (we have typically $R/r_{0}\sim 0.04$, see formula
(\ref{H})) the last term in equation (\ref{motiongeneral}) can be expanded to
first order in the displacement ${\bf r}-{\bf r}_{0}$. This is the so-called
``epicyclic approximation''. Introducing a set of cartesian coordinates $(x,y)$
centered on the vortex and such that the $y$-axis  points in the direction
opposite to the sun, the particle equation (\ref{motiongeneral}) reduces to:
 \begin{equation}
{d^{2}x\over dt^{2}}=-\xi\biggl ({dx\over dt}-u_{x}\biggr ) +2\Omega {dy\over
dt} 
\label{cartx}
\end{equation}
\begin{equation}
{d^{2}y\over dt^{2}}=-\xi \biggl ({dy\over dt}-u_{y}\biggr )-2\Omega {dx\over
dt}+3\Omega^{2}y
\label{carty}
\end{equation}
At sufficiently large distances from the vortex, the velocity field is a simple
shear:
\begin{equation}
u_{x}={3\over 2}\Omega y
\label{Kepshearx}
\end{equation}
\begin{equation}
u_{y}=0
\label{Kepsheary}
\end{equation}
obtained as a first order expansion of the Keplerian velocity around ${\bf
r}_{0}$. Its vorticity is $\omega_{K}\equiv \partial_{x}u_{y}-\partial_{y}
u_{x}=-{3\over 2}\Omega$. 

It remains now to specify the velocity field inside the vortex. We can construct
an exact solution of the incompressible 2D Euler equation by taking an elliptic
patch of uniform vorticity $\omega$. This solution is well-known by model
builders (see, e.g, Saffman 1992) \footnote{This solution was indicated to me by
J.Sommeria.} but, to our knowledge, it has never been applied in an
astrophysical context. Therefore, we give a short description of its
construction in Appendix A. In the vortex, the velocity field writes
 \begin{equation}
u_{x}=-{q^{2}\over 1+q^{2}}\omega y
\label{ux}
\end{equation}
\begin{equation}
u_{y}={1\over 1+q^{2}}\omega x 
\label{uy}
\end{equation}
where $q=a/b$ is the aspect ratio of the elliptic patch ($a,b$ are the semi-axis
in the $x$ and $y$ directions respectively). Outside the vortex, the velocity
field is given by equation  (\ref{psioutellfin}) of Appendix A. At large
distances, we recover the Keplerian shear (\ref{Kepshearx})(\ref{Kepsheary}). 

The matching conditions between the elliptic vortex and the Keplerian shear
require that $\omega$, $\omega_{K}$ and $q$ be related according to (see
Appendix A):
\begin{equation}
{\omega_{K}\over\omega}={q(q-1)\over 1+q^{2}}
\label{relation}
\end{equation} 
The solution (\ref{ux})(\ref{uy}) is valid for $q>1$, implying
$0<\omega_{K}/\omega<1$. The vortex is {\it anticyclonic} ($\omega<0$) and is
oriented with its major axis parallel to the shear streamlines. It can be shown
to be stable to infinitesimal two-dimensional disturbances. For $q\rightarrow 1$
(circular vortices), $\omega\rightarrow -\infty$ and for $q\rightarrow \infty$
(infinitely elongated vortices), $\omega\rightarrow \omega_{K}$. Therefore,
$\omega$ is in the range $\rbrack -\infty,- {3\over 2}\Omega\rbrack$. In a
rotating disk, we expect that $\omega\sim -2\Omega$, corresponding to a Rossby
number of order $1$. This value is achieved by vortices with aspect ratio
$q\simeq 4$, in good agreement with the model of Tanga {\it et al.} (1996). The
epicyclic vortex considered by Barge \& Sommeria (1995) corresponds to $q=2$ and
$\omega=-{5\over 2}\Omega$. These values do not satisfy the matching condition
(\ref{relation}) with the Keplerian shear so the epicyclic vortex is not an
exact solution of the incompressible Euler equation. It may be used, however, as
an approximate solution if we take into account a coupling between the particles
and the gas in the spirit of a two-fluid model.

\subsection{The capture time}
\label{sec_capturetime}

Before going into mathematical details, we recall the argument of Tanga {\it et
al.} (1996) which shows very simply why particles are trapped by anticyclonic
vortices in a rotating disk. Introducing a system of polar coordinates
$(r,\theta)$ whose origin coincides with the vortex center, the radial component
of equation (\ref{motiongeneral}) reads:
\begin{equation}
{d^{2}r\over dt^{2}}= r\biggl ({d\theta\over dt}\biggr )^{2}+2\Omega r
{d\theta\over dt}-\xi\biggl ({dr\over dt}-u_{r}\biggr
)+3\Omega^{2}r\sin^{2}\theta
\label{radial}
\end{equation}
where we have used the epicyclic approximation. The first term is a centrifugal
force due to the rotation of the vortex (not to be confused with the centrifugal
force  $\Omega^{2}r$ due to the rotation of the disk) and the second term is the
Coriolis force. The drag term in equation (\ref{motiongeneral}) forces the
particle velocity to approach the fluid velocity, i.e ${d\theta\over dt}\sim
{\omega}$. For cyclonic vortices ($\omega>0$), both centrifugal and Coriolis
forces are positive and push the particles outward. For anticyclonic vortices
($\omega<0$), the Coriolis force pushes the particles inward and comes in
conflict with the centrifugal force which is always directed outward. If the
vortex rotates rapidely, the centrifugal force prevails over the Coriolis force
and the particle is expelled. In the other case, the Coriolis term dominates and
the particle is captured. We conclude that only slowly rotating anticyclonic
vortices can capture dust particles. This trapping process is specific to
rotating fluids; in ordinary simulations of two-dimensional turbulence (without
Coriolis force) the particles never penetrate the vortices. Note that the
epicyclic acceleration (last term in equation (\ref{radial})) is always directed
outward. This term is responsible for a flux of particles toward the sun (if the
gravitational force dominates, i.e for $y<0$) or toward the outer nebula (if the
centrifugal force dominates, i.e $y>0$). However, this flux is quite small and
doesn't affect the overall trapping process (see Tanga {\it et al.} 1996).

We give now a more precise analysis of the trapping process and try to derive an
explicit expression for the capture time of the particles as a function of their
friction parameter $\xi$. To that purpose, we notice that equations
(\ref{cartx}) (\ref{carty}) with the velocity field (\ref{ux}) (\ref{uy}) form a
{\it linear} system of coupled differential equations. We seek therefore a
solution of the form:
\begin{equation}
x= X e^{\lambda t}
\label{x}
\end{equation}
\begin{equation}
y= Y e^{\lambda t}
\label{y}
\end{equation}
where $X,Y$ and $\lambda$ are complex numbers. Substituting into equations
(\ref{cartx}) (\ref{carty}), we obtain a linear system of algebraic equations:
\begin{equation}
-\lambda(\lambda+\xi)X+2\Omega\biggl ({3\over 4} {q\over q-1}\xi+\lambda\biggr
)Y=0
\label{s1}
\end{equation}
\begin{equation}
\biggl ({3\Omega\over 2}{1\over q(q-1)}\xi+2\Omega\lambda\biggr
)X+(\lambda\xi+\lambda^{2}-3\Omega^{2})Y=0
\label{s2}
\end{equation} 
There are nontrivial solutions only if the determinant of this system is zero.
We are led therefore to a fourth order polynomial equation in $\lambda$:
\begin{eqnarray}
\lambda^{4}+2\xi\lambda^{3}+(\xi^{2}+\Omega^{2})\lambda^{2}+3{1+q\over
q(q-1)}\xi\Omega^{2}\lambda\nonumber\\ 
+{9\over
4}\xi^{2}\Omega^{2}{1\over(q-1)^{2}}=0
\label{lambdaeq}
\end{eqnarray}
By definition, we will say that a particle is {\it light} or {\it heavy} whether
$\xi>\Omega$ or $\xi<\Omega$ respectively. This terminology will take more sense
in the sequel (see in particular section \ref{sec_application1}). We now
consider the asymptotic limits $\xi\rightarrow\infty$ and $\xi\rightarrow 0$ of
equation (\ref{lambdaeq}).

For $\xi\rightarrow\infty$ (light particles), the four roots of equation
(\ref{lambdaeq}) are
\begin{equation}
\lambda=-\xi \qquad {\rm (double \ root)}
\label{ll1}
\end{equation}
\begin{equation}
\lambda=\pm {3\Omega\over 2(q-1)}i-{3\Omega^{2}\over 4\xi}{(q-2)(2q+1)\over
q(q-1)^{2}}
\label{ll2}
\end{equation}
The first solution is rapidely damped and will not be considered anymore. The
second solution describes the rotation of the dust particles in the vortex:
\begin{equation}
x=A\cos\biggl \lbrack -{3\Omega\over 2(q-1)}t\biggr\rbrack e^{-t/t_{capt}}
\label{x2}
\end{equation}
\begin{equation}
y={A\over q}\sin\biggl \lbrack -{3\Omega\over 2(q-1)}t\biggr\rbrack    e^{-
t/t_{capt}}
\label{y2}
\end{equation}
The particles follow ellipses of aspect ratio $q$ and move with angular velocity
$-{3\Omega\over 2(q-1)}$. In fact, for $\xi\rightarrow\infty$, the drag term in
equation (\ref{motiongeneral}) implies ${d{\bf r}\over dt}\simeq {\bf u}$ so, in
a first approximation, the particles just follow the vortex streamlines (their
angular velocity coincides with the angular velocity of the fluid particles, see
Appendix A). In addition, they experience a {\it drift} toward the center due to
the combined effect of the friction and the  Coriolis force. They reach the
vortex center in a typical time:
\begin{equation}
t_{capt}={4\xi\over 3\Omega^{2}}{q (q-1)^{2}\over (q-2)(2q+1)}
\label{tcapt1}
\end{equation} 
defined as the {\it capture time}. Note that for light particles, $t_{capt}$
increases {\it linearly} with $\xi$. 

For $\xi\rightarrow 0$ (heavy particles), the four roots of equation
(\ref{lambdaeq}) are:
\begin{equation}
\lambda=-{3\over 2}{1+q\pm\sqrt{1+2q}\over q(q-1)}\xi
\label{lh1}
\end{equation}
\begin{equation}
\lambda=\pm {\Omega}i-{(q-3)(2 q+1)\over 2 q(q-1)}\xi
\label{lh2}
\end{equation}
The second solution describes again a damped rotation of the particles but with
a different trajectory:
\begin{equation}
x=A\cos(-\Omega t)e^{-t/t_{capt}}
\label{x3}
\end{equation}
\begin{equation}
y={A\over 2}\sin(-\Omega t)e^{- t/t_{capt}}
\label{y3}
\end{equation}
The particles follow ellipses with aspect ratio $2$ and move with angular
velocity $-\Omega$. This is natural since heavy particles have the tendency to
decouple from the gas and reach a pure epicyclic motion. However, due to a
slight friction, they sink progressively in the vortex with a characteristic
time
\begin{equation}
t_{capt}={2 q(q-1)\over (q-3)(2 q+1)}{1\over\xi}
\label{tcapt2}
\end{equation} 
For heavy particles, the capture time increases like $\xi^{-1}$. Very heavy
particles can even leave the vortex. Formally, this possibility is taken into
account in the first solution (\ref{lh1}) which corresponds to open
trajectories. The motion of heavy particles is therefore more complicated and
demands a numerical integration of the particle equation inside and outside the
vortex. This study will not be undertaken in that paper. We shall only consider
the case of closed trajectories described by equations (\ref{x3})(\ref{y3}).

 \begin{figure}
  \includegraphics{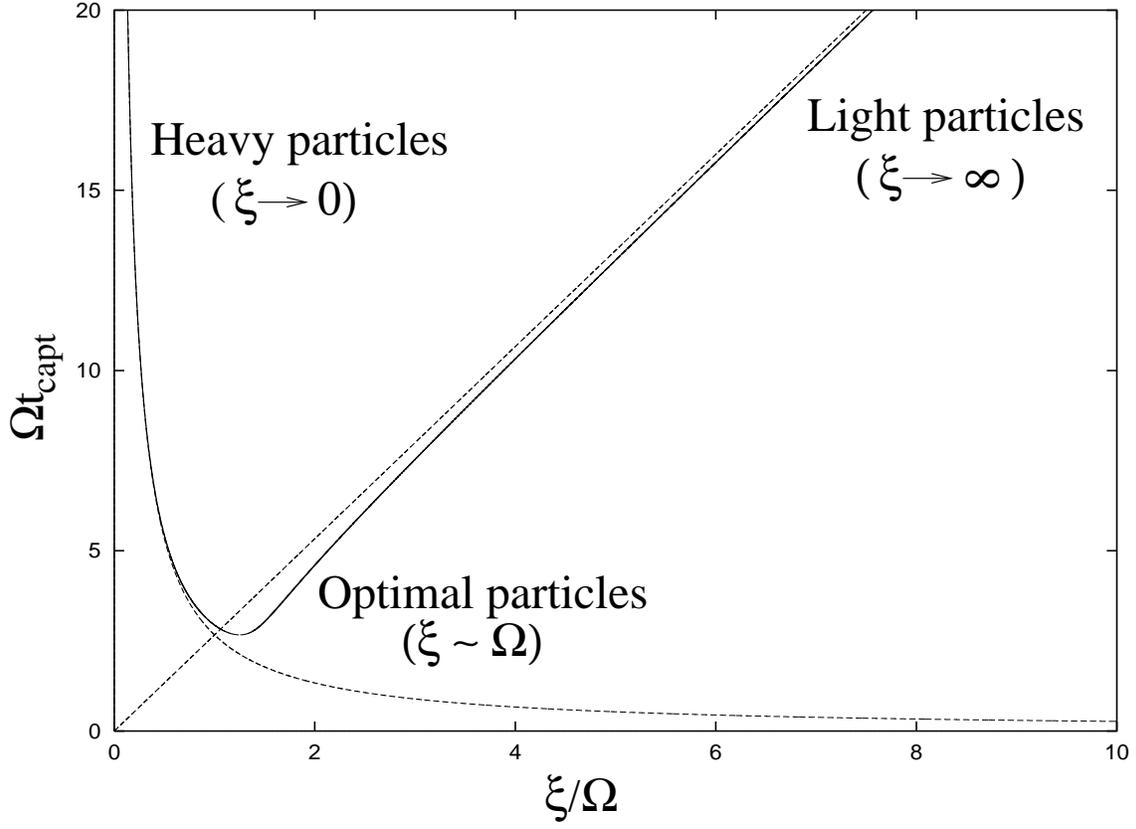}    
\caption[]{Plot of $t_{capt}$ vs $\xi$ for vortices with aspect ratio $q=4$. The
dash lines correspond to formulae (\ref{tcapt1}) and (\ref{tcapt2}) valid for
light ($\xi\rightarrow\infty$) and heavy ($\xi\rightarrow 0$) particles.}
         \label{fig_tcaptvsxi}
   \end{figure}

\begin{figure}
   \includegraphics{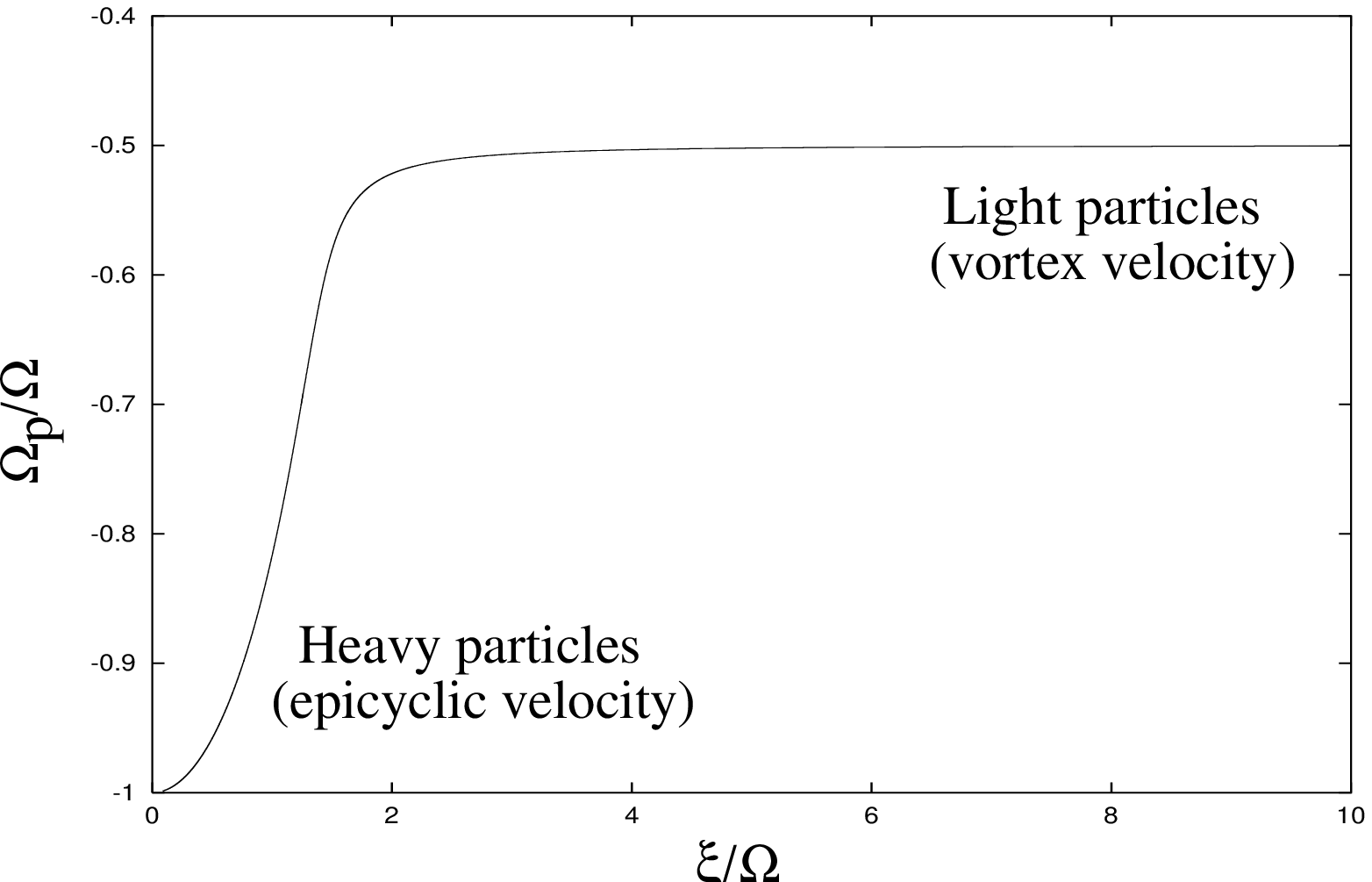}   
\caption[]{Plot of $\Omega_{p}$ (angular velocity of the particles) vs $\xi$ for
vortices with aspect ratio $q=4$ . Light particles ($\xi\rightarrow\infty$) move
with the vortex velocity $-{3\Omega\over 2(q-1)}$ and heavy particles
($\xi\rightarrow 0$) with the epicyclic velocity $-\Omega$.}
              
         \label{fig_omegapvsxi}
   \end{figure}

For intermediate values of $\xi$, the capture time and the angular velocity of
the particles are plotted on figures \ref{fig_tcaptvsxi} and
\ref{fig_omegapvsxi} (for the particular value $q=4$). We see that $t_{capt}$
presents an optimum at $\xi=\xi_{opt}$. Moreover, the asymptotic expressions
(\ref{tcapt1}) and  (\ref{tcapt2}) agree reasonably well with the exact solution
for all the values of the friction parameter. Therefore, considering the
intersection of the asymptotes, we find that the capture time is minimum for 
\begin{equation}
\xi_{opt}\simeq \biggl\lbrack {3(q-2)\over 2(q-1)(q-3)}\biggr\rbrack^{1/2}\Omega
\label{xic}
\end{equation} 
and we have:
\begin{equation}
t^{min}_{capt}\simeq\biggl\lbrack {8(q-1)^{3}q^{2}\over
3(q-3)(2q+1)^{2}(q-2)}\biggr\rbrack ^{1/2}{1\over\Omega} 
\label{tauc}
\end{equation}

According to equations (\ref{tcapt1}) and (\ref{tcapt2}), the condition for
particle trapping ($t_{capt}>0$) requires that $q>3$. This implies that the
vorticity must be in the range [see equation (\ref{relation})]:
\begin{equation}
-{5\over 2}\Omega<\omega<-{3\over 2}\Omega
\label{rangerossby}
\end{equation} 
Particles are ejected from rapidely rotating anticyclones in agreement with the
qualitative discussion of Tanga {\it et al.} (1996) recalled at the begining of
this section.  In the following, we shall specialize on the value of $q$ which
minimizes the capture time $t_{capt}^{min}$. We find $q\simeq 4$. As noted
already, this value corresponds to $\omega\simeq -2\Omega$, i.e a Rossby number
of order $1$. For these vortices, the optimal friction parameter is $\xi_{opt}=
\Omega$ and the corresponding capture time  $t_{capt}^{min}= {8\over 3\Omega}$
is of the order of one rotation period. For $\xi>\Omega$, we have
\begin{equation}
t_{capt}\simeq {8\xi\over 3\Omega^{2}}\qquad ({\rm light \ particles})
\label{taulight}
\end{equation} 
and for $\xi<\Omega$
\begin{equation}
t_{capt}\simeq {8\over 3\xi}\qquad ({\rm heavy \ particles})
\label{tauheavy}
\end{equation} 
Light particles move with angular velocity $-{\Omega\over 2}$ (the vortex
velocity) and heavy particles move with angular velocity $-\Omega$ (the
epicyclic velocity). These results should be compared with the settling time of
the dust particles in the equatorial plane (see, e.g, Nakagawa {\it et al.}
1986). Light particles are settled exponentially with a characteristic time
$\xi/\Omega^{2}$ comparable with equation (\ref{taulight}). On the other hand,
heavy particles undergo overdamped oscillations around the midplane with a
period of the order of $\Omega^{-1}$ and a settling time $2/\xi$ similar to
expression (\ref{tauheavy}). We shall come back on the analogy between the
vortex trapping  and the dust settling in section \ref{sec_diffusion}.

The models of Barge \& Sommeria (1995) and Tanga {\it et al.} (1996) lead to
qualitatively similar results. In the model of Tanga {\it et al.} (1995), the
particles sink to the center of the vortex on a typical time $t_{capt}(\xi)$
which also presents a minimum when $\xi$ is of order $\Omega$. Moreover, for
light particles, $t_{capt}$ increases linearly with $\xi$ like in equation
(\ref{taulight}). The model of Barge \& Sommeria (1995) is physically different
since the particles do not really reach the vortex center but end up on an
epicycle after a time $\sim 1/\xi$. However, the general tendancy is the same.
Light particles ($\xi\gg\Omega$) mainly follow the streamlines of the gas and
stop on epicycles close to the vortex edge (like in case (a) of their figure 1).
Optimal particles with friction parameter $\xi\sim\Omega$ reach deeper
epicycles, almost at the center of the vortex (case (b)).  Heavy particles
($\xi\ll\Omega$) take a long time to connect an epicycle and can even escape
from the vortex since their motion is nearly unaffected by the friction drag.
This is also a possibility in our model and in Tanga {\it et al.} (1996).
Therefore, the three vortex models give relatively similar results even if their
physical contents are  different. The recent numerical simulations of Godon \&
Livio (1999c) for a compressible flow also show a capture optimum when the drag
parameter is of the order of the orbital frequency.

\subsection{The mass capture rate}
\label{sec_capturerate}

In addition to the capture time, an important aspect of the problem concerns the
mass capture rate (Barge \& Sommeria 1995). This quantity can be estimated as
follows. First, we have to determine the capture cross section of the vortex as
a function of the friction parameter $\xi$. Without the effect of the drift, a
particle with impact parameter $\eta$ would cross the vortex in a typical time
$t_{cross}\sim R/u_{K}$, where $u_{K}={3\over 2}\Omega \eta$ is the Keplerian
velocity. The particle will be captured by the vortex if, during that time, the
deflection due to the drift is precisely of order $\eta$. Light particles
($\xi\rightarrow\infty$) follow the edge of the vortex and, for them,
$v_{drift}\sim R/t_{capt}$. On the other hand, heavy particles ($\xi\rightarrow
0$) keep their rectilinear motion and enter directly the vortex so that, for
them,  $v_{drift}\sim \eta/t_{capt}$. The critical parameter $\eta_{c}$ is given
by the condition $t_{cross}\times v_{drift}\sim\eta_{c}$. Moreover, for optimal
particles with $\xi\sim\Omega$, we expect that $\eta_{c}\sim R$. Regrouping all
these results, we obtain    
\begin{equation}
f(\xi)={\eta_{c}\over R}\simeq \biggl ({\Omega\over \xi}\biggr )^{1/2}\qquad
({\rm light \ particles})
\label{f}
\end{equation} 
\begin{equation}
f(\xi)={\eta_{c}\over R}\simeq {\xi\over\Omega}\qquad ({\rm heavy \ particles})
\label{fbis}
\end{equation}
These results agree with the asymptotic behaviour of the function $f(\xi)$
determined numerically by Barge \& Sommeria (1995) in their model.

The mass capture rate can be estimated by assuming that the particles are
carried to the vortex by the Keplerian shear $u_{K}={3\over 2}\Omega y$, and
that they are continuously renewed by the inward drift (directed towards the
sun) associated with the velocity difference $\Delta V$ between gas and
particles (Barge \& Sommeria 1995). Another alternative, consistent with the
simulations of Bracco {\it et al.} (1999), is that the particles are already
concentrated by the turbulent fluctuations that accompany vortex formation in
the early stages of the protoplanetary nebula. It is likely that both mechanisms
come into play in the capture process. Considering the first possibility, which
can be studied analytically, we have (Barge \& Sommeria 1995):
\begin{equation}
{dM\over dt}=2\int_{0}^{\eta_{c}}\sigma_{d}u_{K}dy={3\over 2}\sigma_{d}\Omega
R^{2}f^{2}(\xi)
\label{mrate}
\end{equation} 
where $\sigma_{d}$ is the surface density of the dust particles outside the
vortex. The total mass collected by the vortex during its lifetime is
\begin{equation}
M_{life}={3\over 2}(\Omega t_{life})\sigma_{d}R^{2}f^{2}(\xi)
\label{mlife}
\end{equation}  
The mass capture rate is proportional to the effective surface $\sim
f^{2}(\xi)R^{2}$ of the vortex. As first emphasized by Barge \& Sommeria (1995),
it is maximum for particles with  $\xi\sim \Omega$.

In conclusion, various vortex models and different physical arguments show that
the capture is {\it optimal} for particles whose friction parameter is close to
the disk rotation. We shall now consider the implications of this result on the
structure of the solar nebula.

\subsection{Application to the solar nebula}
\label{sec_application1}

We shall assume that the solid particles are spherical, of radius $a$ and
density $\rho_{s}$. The value of the friction parameter $\xi$ depends whether
the size of the particles is larger or smaller than the mean free path $\lambda$
in the gas (see, e.g, Weidenschilling 1977). When $a<{9\over 4}\lambda$, we are
in the Epstein regime and:
\begin{equation}
\xi={\Omega\sigma_{g}\over 2\rho_{s}a}
\label{xiepstein}
\end{equation} 
where $\sigma_{g}$ is the gas surface density.

When $a>{9\over 4}\lambda$, the situation is more complicated because, in
general, the friction parameter depends on the velocity difference $|{\bf
v}-{\bf u}|$ between the particles and the gas. There is, however, a regime
where $\xi$ is independant of the particle relative velocity. This is the Stokes
regime in which: 
\begin{equation}
\xi={9\sigma_{g}\Omega\lambda\over 8 a^{2}\rho_{s}}
\label{xistokes}
\end{equation} 
This regime is valid when the particle Reynolds number $R_{p}={2a\over
\nu_{g}}|{\bf v}-{\bf u}|$, where $\nu_{g}={1\over 2}c_{s}\lambda$ is the
kinematic viscosity of the gas,  is smaller than $1$. This is the case for light
particles which move typically with the gas velocity ${\bf v}\simeq {\bf u}$.
For heavy particles on the contrary, we have $R_{p}>1$ and, rigorously speaking,
the friction parameter depends on the relative velocity. The use of a more exact
expression for $\xi$ would require a numerical integration of the particle
trajectory but the results shouldn't dramatically differ from those obtained
with expression (\ref{xistokes}). Since we are mainly interested by orders of
magnitude, we will use expression (\ref{xistokes}) for all particle sizes (with
$a>{9\over 4}\lambda$), but keep in mind this uncertainty for large particles.

We shall adopt a standard model of the solar nebula, following Cuzzi {\it et
al.} (1993). It corresponds to a ``minimum mass'' circumstellar nebula with
parameters: 
\begin{equation}
\Omega=2\pi\biggl ({r\over 1{\rm A.U}}\biggr )^{-{3/2}}{\rm years}^{-1}
\label{omega}
\end{equation}
\begin{equation}
\sigma_{d}=10\biggl ({r\over 1 A.U}\biggr )^{-3/2} {\rm g}/{\rm cm}^{2}
\label{sd}
\end{equation}
\begin{equation}
\sigma_{g}=1700\biggl ({r\over 1{\rm A.U}}\biggr )^{-{3/ 2}}{\rm g}/{\rm cm}^{2}
\label{sg}
\end{equation} 
\begin{equation}
H=0.04 \biggl ({r\over 1A.U}\biggr )^{5/4}A.U
\label{H}
\end{equation}
\begin{equation}
\lambda=1 \biggl ({r\over 1{\rm A.U}}\biggr )^{{11/4}}{\rm cm} 
\label{lambda}
\end{equation} 
We take $1\ {\rm year}=3.15\ 10^{7} s$ and $1\ A.U=1.49\ 10^{13} cm$.

For a given type of particles, the transition between the Epstein and the Stokes
regime is achieved at a specific distance from the sun given by: 
\begin{equation}
r_{c}=\biggl ({4\over 9}{a\over 1{\rm cm}}\biggr )^{{4\over 11}}{\rm A.U} 
\label{rc}
\end{equation} 
We are particularly interested by particles of order $10$ cm in size. Indeed,
smaller particles can grow by aggregation processes without the aid of vortices.
However, when they reach decimetric sizes, collisional fragmentation becomes
prohibitive and prevents further evolution (on relevant time scales). It is
therefore at this range of sizes that the vortex scenario should come at work
and facilitate the formation of bigger structures, the so-called planetesimals.
For particles between $10$ and $100$ cm in size, we find that the critical
radius (\ref{rc}) at which the gas drag law changes lies in the range:
\begin{equation}
1.7 A.U <r_{c}< 3.9 A.U
\label{rcrange}
\end{equation} 
that is to say just at the separation between telluric (inner) and giant (outer)
planets. This result is relatively robust because it depends only on a small
power  of the particles size and is independant on their density. We can
therefore wonder if there is not a connection between the division of the solar
system in two groups of planets (telluric and gaseous) and the two regimes
(Stokes and Epstein) of the gas drag law in the primordial nebula. In the
following we show how the vortex scenario can give further support to this idea.

Since the friction coefficient of a particle with size $a$ and density
$\rho_{s}$ depends on parameters (like $\sigma_{g}$, $\lambda$ and $\Omega$)
which are functions of the distance $r$ from the sun, the friction coefficient
$\xi$ is itself a function of $r$. Combining equations (\ref{xiepstein})
(\ref{xistokes}) with equations (\ref{omega}) (\ref{sg}) (\ref{lambda}), we
obtain:
\begin{equation}
{\xi\over\Omega}={1913\over a^{2}\rho_{s}}r^{5\over 4} \quad {\rm if}\quad
r<r_{c}\quad ({\rm Stokes \ zone})
\label{xiomstokes}
\end{equation} 
\begin{equation}
{\xi\over\Omega}={850\over a\rho_{s}}r^{-{3\over 2}} \quad {\rm if}\quad
r>r_{c}\quad ({\rm Epstein \ zone})
\label{xiomepstein}
\end{equation}
where $r$ is measured in A.U, $a$ in cm and $\rho_{s}$ in g/cm$^{3}$.

\begin{figure}
    \includegraphics{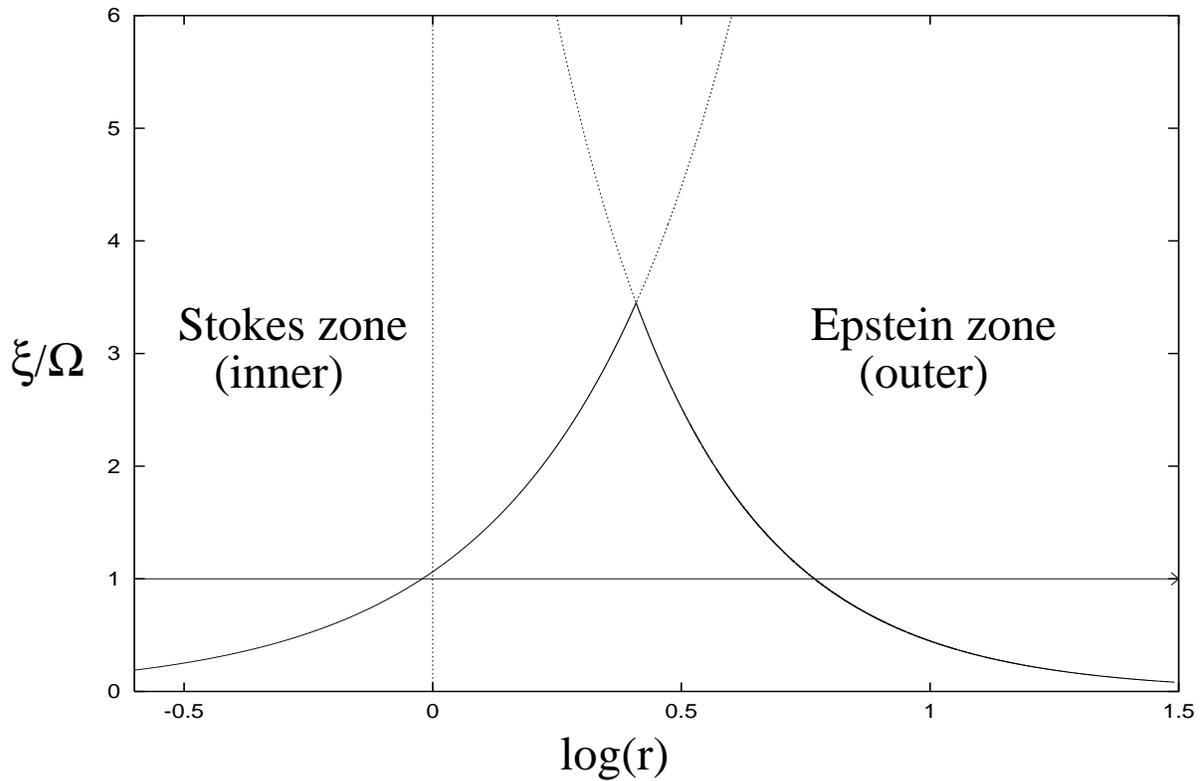}  
\caption[]{Evolution of the friction parameter $\xi/\Omega$ as a function of the
distance to the sun for a given type of particles (the figure corresponds to
particles with size  $a=30\ cm$ and bulk density $\rho_{s}=2 \ g/cm^{3}$). The
friction parameter is maximum at $r=r_{c}$ where the gas drag law passes from
the Stokes to the Epstein regime. The condition $\xi/\Omega=1$ determines two
optimal regions in the disk.}
              
         \label{fig_xivslogr}
   \end{figure}

\begin{figure}
    \includegraphics{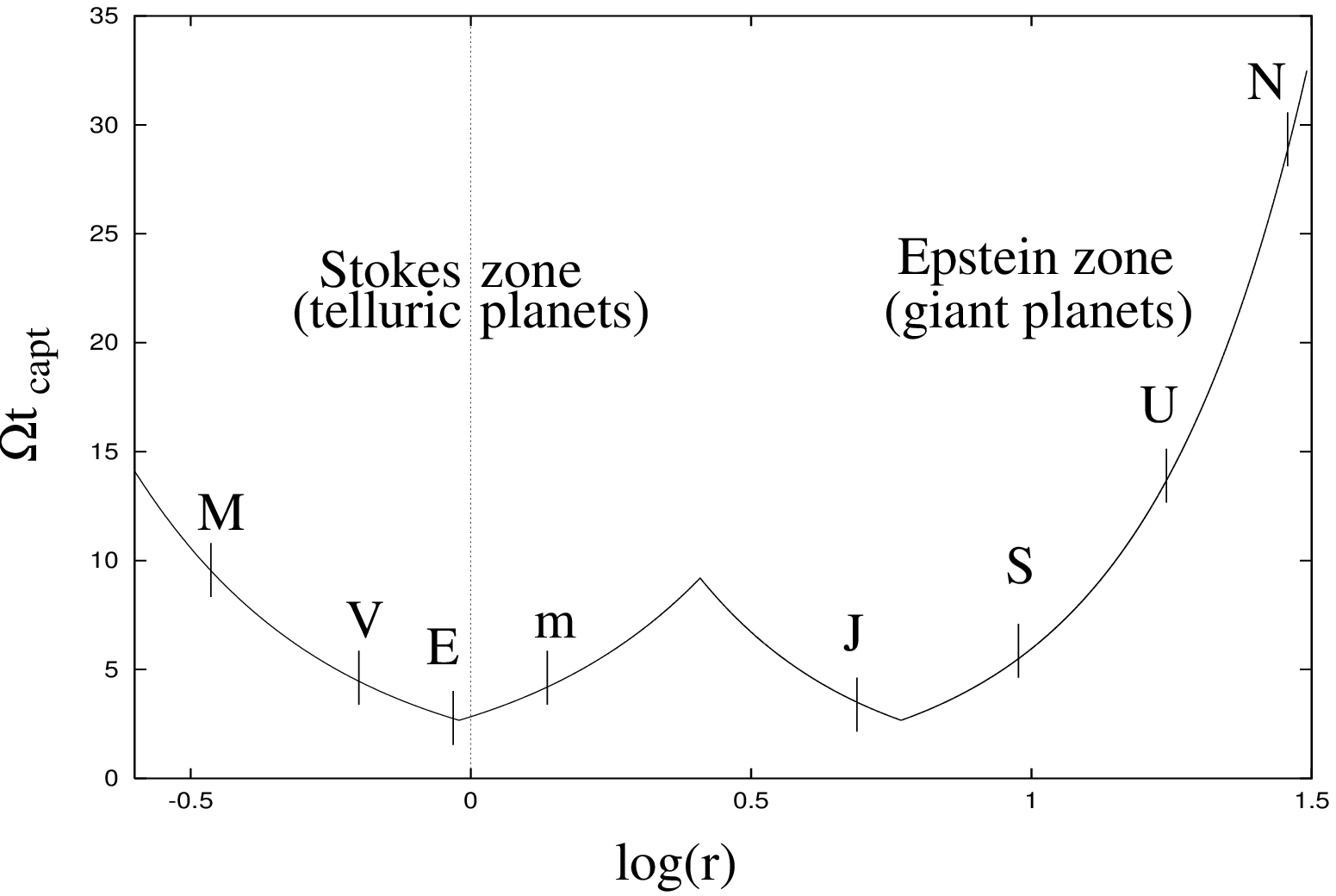}   
\caption[]{Evolution of the capture time $\Omega t_{capt}$ as a function of the
distance to the sun for a given type of particles ($a=30\ cm$, $\rho_{s}=2 \
g/cm^{3}$). We have represented the present position of the planets. The capture
time is optimum near the Earth (in the Stokes zone) and between Jupiter and
Saturn (in the Epstein zone).}
              
         \label{fig_tcaptvslogr}
   \end{figure}

According to the results of section \ref{sec_capturetime}, the capture time
$\Omega t_{capt}$ (measured in rotation periods) is optimal when $\xi/\Omega=1$.
Since $\xi/\Omega$ is a function of $r$ with a maximum at $r_{c}$, this
criterion  determines {\it two} prefered locations in the disk, one in each zone
(see figure \ref{fig_xivslogr}). In the Stokes (inner) zone, the optimum is at:
\begin{equation}
r_{in}=\biggl ({a^{2}\rho_{s}\over 1913}\biggr )^{4/5}
\label{rin}
\end{equation}
and in the Epstein (outer) zone, it is at:
\begin{equation}
r_{out}=\biggl ({850\over a\rho_{s}}\biggr )^{2/3}
\label{rout}
\end{equation}
More generally, we can combine equations (\ref{taulight}) (\ref{tauheavy}) with
equations (\ref{xiomstokes}) (\ref{xiomepstein}) to express the capture  time
$\Omega t_{capt}$ as a function of the distance to the sun (for a given type of
particles). We find a W-shaped curve (see figure \ref{fig_tcaptvslogr})
determined by the power laws:
\begin{equation}
\Omega t_{capt}={a^{2}\rho_{s}\over 717} r^{-5/4}\qquad (r<r_{in})
\label{z1}
\end{equation}
\begin{equation}
\Omega t_{capt}={5101\over a^{2}\rho_{s}}r^{5/4}\qquad (r_{in}<r<r_{c})
\label{z2}
\end{equation}
\begin{equation}
\Omega t_{capt}={2267\over a\rho_{s}}r^{-3/2}\qquad (r_{c}<r<r_{out})
\label{z3}
\end{equation}
\begin{equation}
\Omega t_{capt}={a\rho_{s}\over 319}r^{3/2}\qquad (r>r_{out})
\label{z4}
\end{equation}

In order to make a numerical application, we assume that all the particles have
the same density $\rho_{s}\sim 2 g/cm^{3}$ (the density of a composite rock-ice
material) and the same size $a\sim 30 cm$ (a typical prediction of grain growth
models). In principle, we should consider a spectrum of sizes and densities but
we choose these particular values in order to illustrate at best the predictions
of the vortex model. For these values, the optimum in the inner zone is at
$r_{in}\sim 1 A.U$, i.e near the Earth orbit, and the optimum in the outer zone
is at $r_{out}\sim 6 A.U$, i.e  between Jupiter and Saturn's orbits. The
transition between the Stokes and the Epstein regime occurs at $r_{c}\sim 2.7
A.U$, i.e near the asteroid belt.

 \begin{table}
\caption[]{Minimum sizes of dust particles (in $cm$) which can trigger the
gravitational instability ($\rho_{s}=2g/cm^{3}$). }
         \label{table}
      \[
         \begin{array}{l l l l l l l }
            \hline
            \noalign{\smallskip}
Planets    & r (A.U) & \lambda & a_{opt}  & a_{min}^{conc} & a_{min}^{capt} &
a_{min} \\
            \noalign{\smallskip}
            \hline
            \noalign{\smallskip}
	   {\rm Telluric \ planets:} &   &  &  & & &   \\
            Mercury & 0.387  & 0.07 & 17 &14 &7 &3   \\
            Venus & 0.723 & 0.4 & 25 &20  &11  &5 \\
	    Earth & 1 & 1 & 31 &25  &13  &6 \\
	    Mars & 1.52 &3  &40  &33  &17  &8 \\
         \hline
           {\rm Asteroid \ belt:} & 3  & 20 &61  &50 &15 &3   \\
          \hline
	   {\rm Giant \ planets:} &  &  &  & & &   \\
	    Jupiter & 5.20 &93  &36  &23  &7  &1 \\
            Saturn & 9.52 &492  &14  & 9 &3  &0.5 \\
	    Uranus & 19.2 &3364  &5  &3  &1  &0.2 \\
            Neptune & 30.0 &11523  &3  &2  &0.5  &0.1 \\
           \noalign{\smallskip}
            \hline
         \end{array}
     \]
   \end{table}

\begin{figure}
     \includegraphics{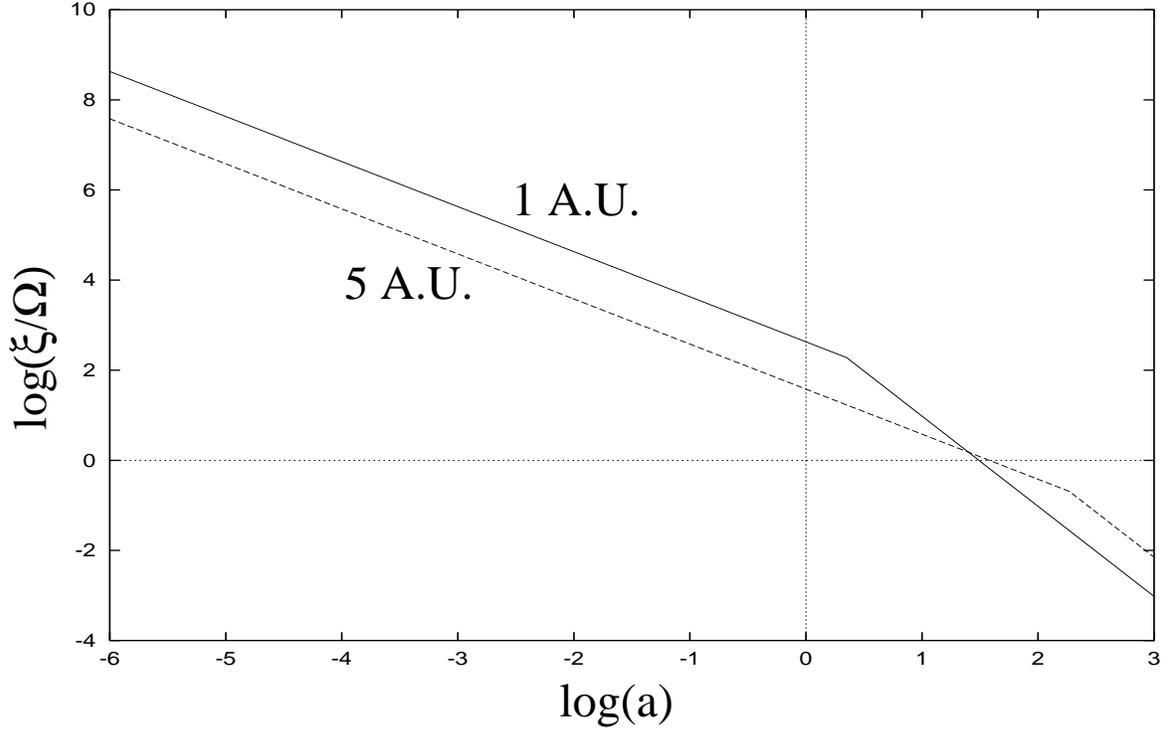}    
\caption[]{Variation of the friction parameter with the size of the particles at
$r=1\ A.U$ and $r=5\ A.U$ (we have taken $\rho_{s}=2 g/cm^{3}$). The capture is
optimum for particles with friction parameter $\xi\sim\Omega$. This corresponds
to decimetric sizes in the region of the planets. }
         \label{fig_xivsa}
   \end{figure}

\begin{figure}
     \includegraphics{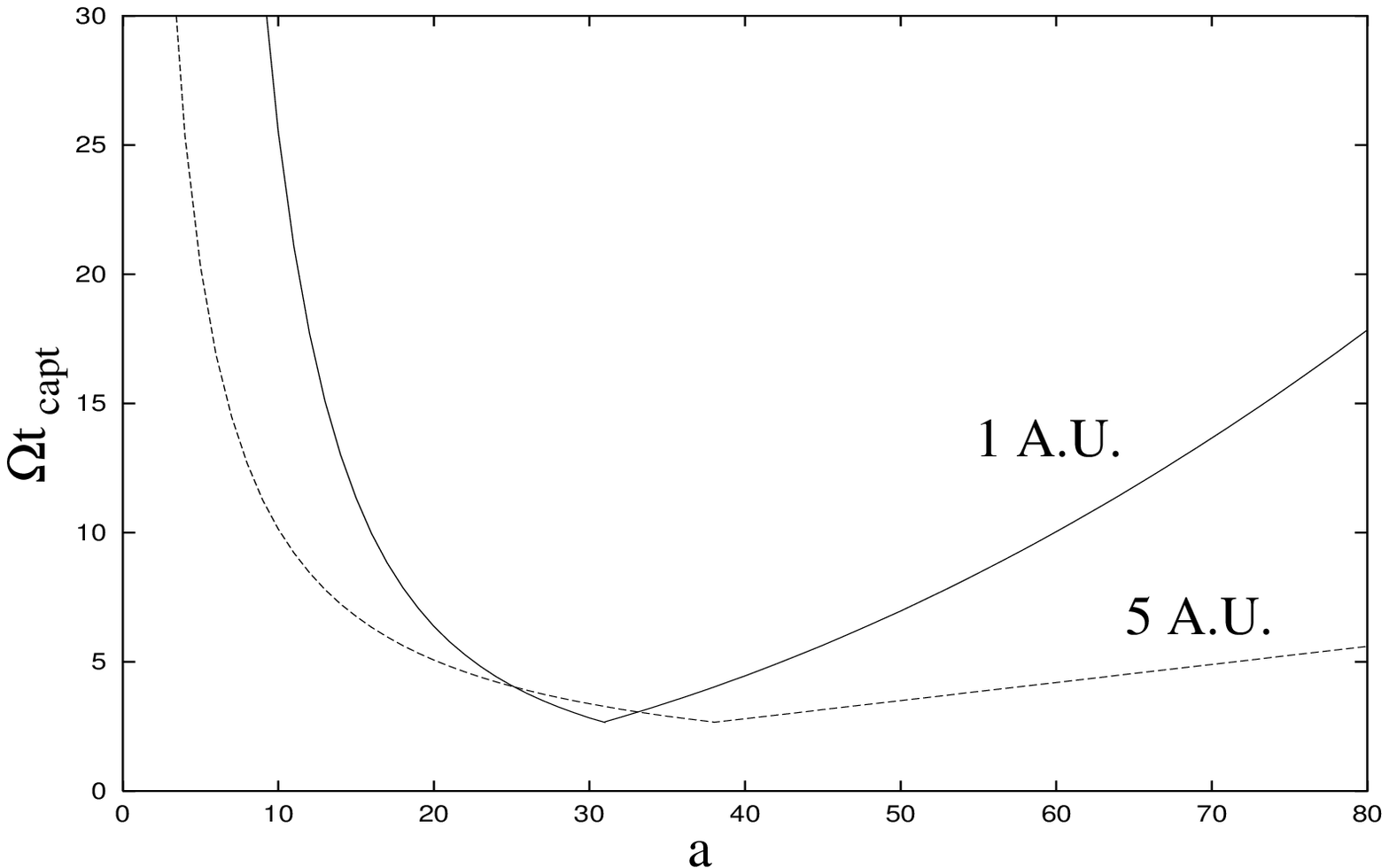}    
\caption[]{Variation of the capture time with the size of the the particles at
$r=1\ A.U$ and $r=5\ A.U$ ($\rho_{s}=2 g/cm^{3}$). The capture time is minimum
for particles with radius $a\sim 30cm$.  }
         \label{fig_tcaptvsa}
   \end{figure}

Alternatively,  we can determine, at each heliocentric distance, the size of the
particles which are preferentially concentrated by the vortices. The results are
indicated on table \ref{table} (see also figures \ref{fig_xivsa} and
\ref{fig_tcaptvsa}). They show that the optimal sizes lie between $1$ and $50$
cm in the region of the planets (for a bulk density $\rho_{s}=2 g/cm^{3}$). Such
particles are concentrated in the vortices after only one rotation period. By
contrast, the capture time for particles of $1\mu m$ in size (the initial size
of the dust grains in the primordial nebula) is $\Omega t_{capt}\sim 10^{9}$ (we
have a similar characteristic value for their settling time). This exceeds the
lifetime of a circumstellar disk by many orders of magnitude. These results (see
also sections \ref{sec_application2} and \ref{sec_application3}) indicate that
sticking of particles up to cm-sized bodies is an indispensable step in the
process of planet formation.

\begin{figure}
      \includegraphics{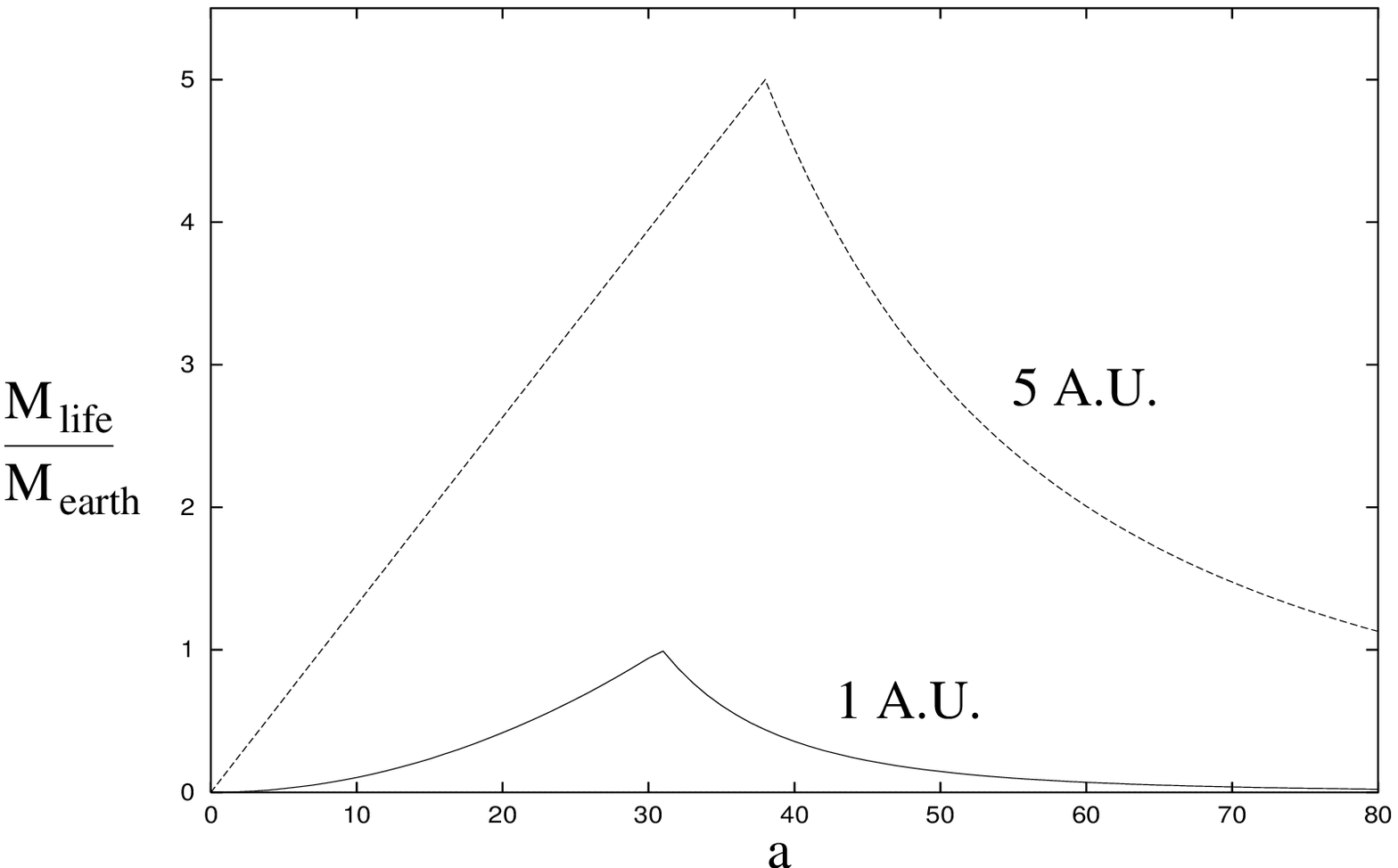}  
\caption[]{Variation of $M_{life}$ with the size of the particles at $r=1\ A.U$
and $r=5\ A.U$ ($\rho_{s}=2 g/cm^{3}$).  }
         \label{fig_mlifevsa}
   \end{figure}

\begin{figure}
     \includegraphics{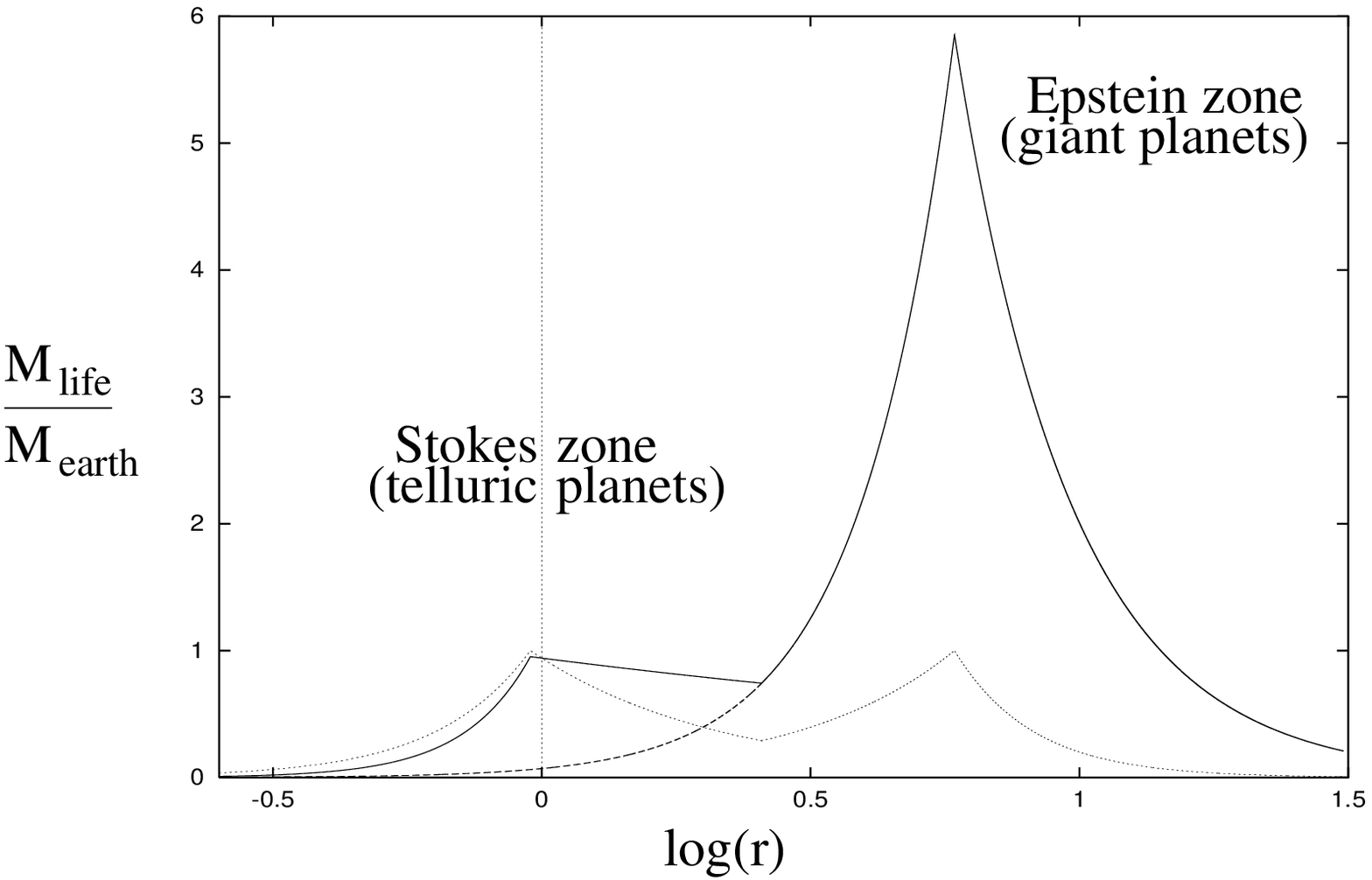}    
\caption[]{Variation of $M_{life}$ with the heliocentric distance for particles
with size $a=30cm$ and bulk density  $\rho_{s}=2 g/cm^{3}$.  }
         \label{fig_mlifevsr}
   \end{figure}

We can also study how the mass captured by the vortices varies throughout the
nebula and how it depends on the size of the particles. According to equations
(\ref{mlife}) (\ref{sd}) and (\ref{H}), the mass captured by a vortex during its
lifetime can be written:
\begin{equation}
{M_{life}\over M_{\oplus}}=8.916\ 10^{-4}(\Omega t_{life}) r f^{2}(\xi)
\label{Mlifenorm}
\end{equation}
where we have introduced the Earth mass $M_{\oplus}=5.976\ 10^{27} \ g$ as a
normalization factor. At $r=1 A.U$ and for optimal particles with size $a\sim 30
cm$ and bulk density $\rho_{s}\sim 2 g/cm^{3}$ (see table \ref{table}), we have
$\xi\sim\Omega$ leading to $r f^{2}\sim 1$. Since the vortex lifetime is not
accurately known, it makes sense to calibrate its value so as to satisfy
$M_{life}(1A.U)=M_{\oplus}$. This yields $\Omega t_{life}\simeq 1122$,
corresponding to $\sim 180$ rotation periods, in agreement with the estimate of
Barge \& Sommeria (1995) based on the value of the disk $\alpha$-viscosity and
with the numerical simulations of Godon \& Livio (1999a,b,c). Figure
\ref{fig_mlifevsa} shows how the  mass $M_{life}$ depends on the size of the
particles at a given position of the solar nebula. We can also determine how it
varies with the heliocentric distance for a given type of particles. The results
are reported on figure \ref{fig_mlifevsr} (full line) for particles with $a\sim
30 cm$ and $\rho_{s}\sim 2 g/cm^{3}$. The captured mass presents a global
maximum near Jupiter's orbit and a plateau in the Earth's region. These results
complete the work of Barge \& Sommeria (1995) who first noticed the existence of
an optimum near Jupiter. However, they didn't consider the Stokes regime in
their article and made the wrong statement that ``inside Jupiter's orbit,
particle concentration occurs in an annular region at the vortex periphery
[because the friction parameter $\xi$ is larger]''. In reality, the friction
parameter decreases anew when we enter the Stokes domain at $r<r_{c}$ (see
figure \ref{fig_xivslogr}). This allows the existence of {\it another} optimum
near the Earth orbit. Therefore, the mass collected by the vortices in the inner
zone is larger that one would obtain without this change of regime (the dash
line in figure \ref{fig_mlifevsr} would be found if the Epstein law was used in
the whole disk). This transition may explain  the division of the solar system
in two groups of planets. Moreover, the vortex scenario explains naturally the
disymmetry between these two groups: the mass capture rate is larger in the
outer zone simply because the vortices are larger. Indeed, the captured mass is
not only proportional to $f^{2}$ (which would give the symmetrical dotted line
of figure \ref{fig_mlifevsr}) but also to the product $\sigma_{d}R^{2}$ which
increases linearly with $r$. This effect may explain the difference in size and
mass between telluric and giant planets. Moreover, the mass captured by the
vortices in the outer zone should be larger than these estimates since they
intercept in priority the matter drifting towards the sun. On the other hand,
the intermediate region at $\sim 3\ A.U$ should be further depleted due to its
proximity with the global maximum.

In conclusion, we find that there are two locations in the primordial nebula
where the trapping of dust by vortices is optimal. These locations belong to the
Stokes and to the Epstein zones are fall near the Earth and Jupiter's orbit
respectively. The zone of transition of the gas drag law is consistent with the
position of the asteroid belt which marks the separation between telluric and
giant planets. The asymmetry between the two groups of planets may be related to
the size of the vortices which are bigger in the outer zone and therefore
capture more mass. The exact values of $r_{in}$, $r_{c}$ and $r_{out}$ depend on
the spectrum of size of the particles, which is not well-known, but the W-shape
of figures \ref{fig_tcaptvslogr} and \ref{fig_mlifevsr} is generic and agrees
with the global structure of the solar system.

\section{Stochastic motion of a particle in a vortex}
\label{sec_stochastic}

\subsection{The diffusion equation}
\label{sec_diffusion}

Due to small-scale turbulence, the motion of a particle in a vortex is not
deterministic but {\it stochastic}. Turbulent fluctuations produce some kicks
which progressively deviate the particle from its unperturbed trajectory. This
is similar to what happens to a colloidal particle in suspension in a liquid
(Brownian motion). An individual fluctuation has a minute effect on the  motion
of the particle, but the repeated action of these fluctuations gives rise to a
macroscopic process of diffusion. The effect of turbulence on the sedimentation
of dust particles in the protoplanetary nebula has been considered in detail by
Dubrulle {\it et al.} (1995). We use a similar approach to study the effect of
turbulence on the capture of dust by large-scale vortices.

The transport equation governing the evolution of the dust surface density
inside a vortex can be written:
\begin{equation}
{\partial \sigma_{d}\over\partial t}+\nabla (\sigma_{d}\langle {\bf
v}\rangle)=\nabla(D\nabla \sigma_{d})
\label{transport}
\end{equation} 
where $\langle {\bf v}\rangle$ is the mean velocity of the particles and $D$
their diffusivity. The mean velocity $\langle {\bf v}\rangle$ is given by the
deterministic model of section \ref{sec_capturetime}. According to equations
(\ref{x2})(\ref{y2}) or (\ref{x3}) (\ref{y3}), it can be written: 
\begin{equation}
\langle {\bf v}\rangle={\bf V}-{1\over t_{capt}}{\bf r}
\label{vaverage}
\end{equation} 
where ${\bf V}$ corresponds to the pure rotation of the particles in the vortex
and $-{\bf r}/t_{capt}$ is their drift towards the center. In the case of light
particles ($\xi\gg\Omega$), ${\bf V}$ is equal to the velocity of the vortex
while in the case of heavy particles ($\xi\ll\Omega$), ${\bf V}$ is equal to the
epicyclic velocity (see section \ref{sec_capturetime}). When (\ref{vaverage}) is
substituted into (\ref{transport}), the diffusion equation takes the form:
\begin{equation}
{\partial \sigma_{d}\over\partial t}+\nabla  (\sigma_{d} {\bf V})=\nabla 
\biggl (D\nabla \sigma_{d}+{\sigma_{d}\over t_{capt}}{\bf r}\biggr )
\label{diffeq}
\end{equation}    
The first term in the r.h.s is a pure diffusion due to small-scale turbulence
and the second term is a drift toward the vortex center due to the combined
effect of the Coriolis force and the vortex (anticyclonic) rotation. Equation
(\ref{diffeq}) illustrates the bimodal nature of turbulence: the diffusion is
due to small-scale fluctuations hardly affected by the rotation of the disk (3D
turbulence) and the trapping process is a consequence of the Coriolis force and
the existence of coherent structures in the disk (2D turbulence).

Since we are mainly interested by orders of magnitude and in order to avoid
unnecessary mathematical complications, we shall consider from now on that the
vortices are circular with typical radius $R$. In this approximation, we can
restrict ourselves to axisymmetric solutions for which the advection term in
equation (\ref{diffeq}) cancels out. We are led therefore to study the
Fokker-Planck equation:
\begin{equation}
{\partial \sigma_{d}\over\partial t}=\nabla 
\biggl (D\nabla \sigma_{d}+{\sigma_{d}\over t_{capt}}{\bf r}\biggr )
\label{fp}
\end{equation} 
For an initial condition consisting of a Dirac function centered at ${\bf
r}_{0}$, there is a well-known analytical solution of equation (\ref{fp}):
\begin{equation}
\sigma_{d}={M\over 2\pi D t_{capt}(1-e^{-2t/t_{capt}})}e^{-{({\bf r}-{\bf
r}_{0}e^{-t/t_{capt}})^{2}\over 2 D t_{capt}(1-e^{-2t/t_{capt}})}}
\label{solution}
\end{equation}
where $M$ is the total mass of particles contained in the vortex. This formula
shows that the relaxation time is equal to the capture time $t_{capt}$ not
affected by turbulence.

The equilibrium solution of the Fokker-Planck equation (\ref{fp}) satisfies the
condition:
\begin{equation}
D\nabla \sigma_{d}+{\sigma_{d}\over t_{capt}}{\bf r}=0
\label{equilibre}
\end{equation}
expressing the balance between the diffusion and the drift. It corresponds to a
gaussian density profile of the form:
\begin{equation}
\sigma_{d}=\sigma_{d}(0)e^{-{r^{2}\over 2D t_{capt}}}
\label{gdensity}
\end{equation} 
This is also the solution of equation (\ref{solution}) for $t\rightarrow
+\infty$. In practice, the equilibrium distribution (\ref{gdensity}) is
established for $t\sim t_{capt}$. Then, the particles are concentrated in the
vortices on a typical length: 
\begin{equation}
l_{d}=\sqrt{2D t_{capt}}
\label{ld}
\end{equation} 

\subsection{Vertical sedimentation}
\label{sec_sedimentation}

Equation (\ref{fp}) is similar to the diffusion equation
\begin{equation}
{\partial n_{d}\over\partial t}={\partial\over\partial z}\biggl ( D{\partial
n_{d}\over\partial z}+{\Omega^{2}\over\xi} n_{d}z\biggr )
\label{dub}
\end{equation} 
used by Dubrulle {\it et al.} (1995) to describe the sedimentation of the dust
particles and determine the sub-disk scale height (see also Weidenschilling
1980). The drift towards the vortex center is replaced in their study by the
drift $-{\Omega^{2}\over \xi}z$ towards the ecliptic plane due to gravity. For
light particles, $t_{capt}\sim {8\xi\over 3\Omega^{2}}$ [see equation
(\ref{taulight})] and the two equations coincide up to numerical factors. This
implies that the particles are concentrated in the vortices on a lenght $l_{d}$
comparable with the sub-disk scale height $H_{d}$ determined by Dubrulle {\it et
al.} (1995) [see their section 3].  

It is relatively straightforward to include the vertical sedimentation of
particles in our study, although we shall specialize in the following on their
{\it horizontal} accumulation in vortices. Introducing the volume density
$\rho_{d}(r,z,t)={1\over H_{d}}n_{d}(z,t)\sigma_{d}(r,t)$ and using equations
(\ref{fp}) and (\ref{dub}), we obtain:
\begin{equation}
{\partial \rho_{d}\over\partial t}=\nabla \biggl \lbrace D \nabla
\rho_{d}+{\Omega^{2}\over\xi} \rho_{d} \biggl ({3\over 8}{\bf r}_{\perp}+{\bf
z}\biggr )\biggr \rbrace
\label{gen}
\end{equation}     
where ${\bf r}_{\perp}$ and ${\bf z}$ are the component of ${\bf r}$ in the
directions perpendicular and parallel to the disk rotation vector ${\mb\Omega}$.
Integrating equation (\ref{gen}) on the vertical direction returns equation
(\ref{fp}) for the surface density.

\subsection{Diffusivity of dust particles}
\label{sec_diffusivity}

It remains now to specify the value of the diffusion coefficient appearing in
equation (\ref{fp}). In general, the turbulent viscosity of the gas is written
under the form $\nu\sim\alpha {c_{s}^{2}\over\Omega}$ where $c_{s}$ is the sound
speed and $\alpha$ a non dimensional parameter which measures the efficiency of
turbulence (Shakura \& Sunyaev 1973). Following current nebula models,
$10^{-4}<\alpha<10^{-2}$. When the turbulence is generated by differential
rotation, $\alpha=2\ 10^{-3}$ (Dubrulle 1992) and we shall take this value for
numerical applications. Since the disk height is $H\sim c_{s}/\Omega$ (see
equation (\ref{Hnebula})), the turbulent viscosity can be written $\nu\sim
\alpha H^{2}\Omega$ or, alternatively, $\nu\sim\alpha\Omega R^{2}$ where $R$ is
the vortex radius.  More precisely, assuming a power law spectrum $E(k)\sim
k^{-\gamma}$ for the gas turbulence (with $\gamma\simeq 5/3$), we have
\begin{equation}
\nu={\alpha\Omega R^{2}\over \sqrt{\gamma+1}}
\label{nu}
\end{equation}
According to Dubrulle {\it et al.} (1995), the diffusivity of the particles can
be written:
\begin{equation}
D=g(\xi)\nu
\label{Dg}
\end{equation}
where
\begin{equation}
g(\xi)=\biggl ({\xi\over\xi+\Omega}{Arctan(B_{k_{0}})\over B_{k_{0}}}\biggr
)^{1/2} 
\label{g}
\end{equation} 
is a function of the friction parameter. The reduction factor of V\"olk {\it et
al.} (1980): 
\begin{equation}
B_{k_{0}}={k_{0}v_{s}\over \xi+\Omega} 
\label{B}
\end{equation} 
depends on the size $k_{0}^{-1}\sim \sqrt{\alpha}H$ of the largest eddies of
turbulence and on the systematic velocity $v_{s}$ of the dust grains. In the
vortices, $v_{s}$ is equal to the drift velocity $r/t_{capt}$.

For light particles ($\xi\gg\Omega$), $g(\xi)\rightarrow 1$ and
\begin{equation}
D={\alpha\Omega R^{2}\over\sqrt{\gamma+1}} \quad ({\rm light \ particles})
\label{Dlight}
\end{equation} 
This result is expectable since light particles mainly follow the streamlines of
the gas. Consequently, their diffusivity tends to the gas turbulent viscosity
$\nu$. 

For heavy particles ($\xi\ll\Omega$), $g(\xi)\rightarrow \sqrt{\xi/\Omega}$ and
\begin{equation}
D={\alpha\Omega R^{2}\over\sqrt{\gamma+1} }\biggl ({\xi\over \Omega}\biggr
)^{1/2}\quad ({\rm heavy \ particles})
\label{Dheavy}
\end{equation}
For $\xi\rightarrow 0$, $D\rightarrow 0$ since there is no coupling with the
gas. For $\xi\sim\Omega$, equations (\ref{Dlight}) and (\ref{Dheavy}) give the
same result, so we can use these expressions in the whole range of friction
parameters. Note, however, that the diffusion approximation is not strictly
valid for heavy particles, so equation (\ref{Dheavy}) must be taken with care. 

Substituting equations (\ref{taulight})(\ref{tauheavy}) and
(\ref{Dlight})(\ref{Dheavy}) into equation (\ref{ld}), the concentration length
can be written explicitely:
\begin{equation}
{l_{d}\over R}=\biggl ({16\alpha\over3\sqrt{\gamma+1}}\biggr )^{1/2}\biggl
({\xi\over\Omega}\biggr )^{1/2}\qquad ({\rm light\  particles})
\label{lRlight}
\end{equation}
\begin{equation}
{l_{d}\over R}=\biggl ({16\alpha\over 3\sqrt{\gamma+1}}\biggr )^{1/2}\biggl
({\Omega\over\xi}\biggr )^{1/4} \qquad ({\rm heavy\  particles})
\label{lRheavy}
\end{equation}
As expected, the concentration is optimum for $\xi\sim\Omega$. In that case, the
particles are distributed over a length $l_{d}\sim\sqrt{\alpha}R$. Lighter and
heavier particles are less concentrated. When  
\begin{equation}
{\xi\over\Omega}> {3\sqrt{\gamma+1}\over 16\alpha}\simeq 153  
\label{range1}
\end{equation}
or
\begin{equation}
{\xi\over\Omega}< \biggl ({16\alpha\over 3\sqrt{\gamma+1}}\biggr )^{2}\simeq 4\
10^{-5} 
\label{range1bis}
\end{equation}
the drift is negligible and there is no concentration ($l_{d}>R$).

\subsection{Application to the solar nebula}
\label{sec_application2}

We now apply these results to the case of the solar nebula and show how the
vortex scenario can make possible the formation of planetesimals at certain
prefered locations of the disk. 

It is generally beleived that planetesimals formed by the gravitational
instability of the particle sublayer (Safronov 1969, Goldreich \& Ward 1973).
The dispersion relation for an infinite uniformly rotating sheet of gas is (see,
e.g, Binney \& Tremaine 1987):
\begin{equation}
\omega^{2}=k^{2}c_{d}^{2}+4\Omega^{2}-2\pi G\sigma_{d}k
\label{dispersion}
\end{equation} 
where $\sigma_{d}$ is the unperturbed surface density of the particles and
$c_{d}$ their velocity dispersion. The particle sub-layer (which behaves as a
very compressible fluid) will be unstable provided that $\omega^{2}<0$, for some
$k$. From equation (\ref{dispersion}), we obtain the criterion for gravitational
instability (Toomre, 1964):
\begin{equation}
c_{d}<{\pi G\sigma_{d}\over 2\Omega}=V_{crit}  
\label{jeans1}
\end{equation}

Once instability is triggered, the system crumbles into numerous planetesimals
of order $10$ km in size. Moreover, the growth time of density perturbation is
predicted to be short, of the order of an orbital period. In addition, the
instability criterion gives the impression that its operation does not require
any sticking mechanism. Goldreich \& Ward (1973) state that ``...the fate of
planetary accretion no longer appears to hinge on the stickiness of the surface
of dust particles''. This is very attractive because sticking mechanisms are
relatively {\it ad hoc} and ill-understood. For these reasons (and also for a
lack of alternatives), the Safronov-Goldreich-Ward scenario was nearly
universaly accepted as the key mechanism for forming planetesimals. Therefore, a
swarm of bodies of a few km in diameter was a common starting point for
numerical simulations of planetary formation.

However, Weidenschilling (1980), followed by Cuzzi {\it et al.} (1993) and
Dubrulle {\it et al.} (1995), realized that this simplisitic picture was ruled
out if the primordial nebula was turbulent. Indeed, turbulence reduces
considerably the vertical sedimentation of the dust particles and prevents
gravitational instability. According to equations (\ref{omega}) and (\ref{sd}),
the velocity threshold  imposed by the instability criterion (\ref{jeans1}) is
of order $V_{crit}=5 cm/s$ throughout the nebula. Even if the nebula as a whole
was perfectly laminar, the formation of a dense layer of particles (considered
as a heavy fluid) would create a turbulent shear with the overlying gas (see,
e.g., Weidenschilling \& Cuzzi 1993). The velocity dispersion of the particles
can be estimated by $c_{d}\sim \sqrt{\alpha}c_{s}$ and easily reaches several
{\it meters} per second (when a numerical value is needed, we shall take
$c_{d}=5 m/s$). Therefore,  the instability criterion (\ref{jeans1}) is not
satisfied (see appendix B for a rigorous derivation  of this criterion in a
turbulent disk). Turbulence is responsible for too high velocity dispersions or,
alternatively, the surface density of the dust sublayer is not sufficient to
trigger the gravitational instability. The density needs to be increased by a
factor $A_{c}\sim 100$ or more to overcome the threshold imposed by Toomre
instability criterion. 

\begin{figure}
     \includegraphics{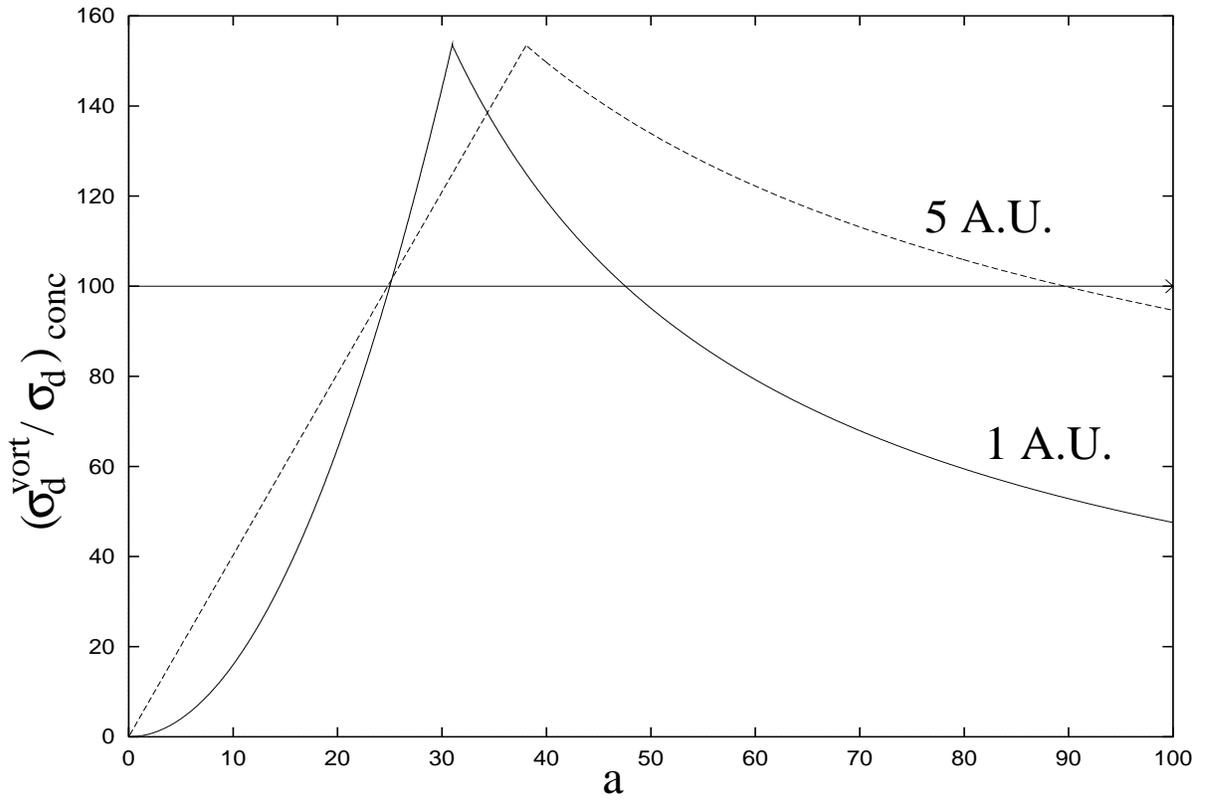}   
\caption[]{Enhancement of the dust surface density inside the vortices (due to
concentration) as a function of the size of the particles at $r=1A.U$ and
$r=5A.U$ ($\rho_{s}=2 g/cm^{3}$).  }
         \label{fig_ampconcvsa}
   \end{figure}

There is therefore a major problem to form planetesimals by gravitational
instability in a turbulent disk. As first suggested by Barge \& Sommeria (1995),
the presence of vortices in the disk can solve this problem. Indeed, by
capturing and concentrating the particles, the vortices can increase locally the
surface density of the dust sublayer and initiate the gravitational instability.
Let us first discuss the {\it concentration} effect. Inside a vortex, an initial
mass of order $\sigma_{d}\pi R^{2}$ is concentrated on a typical length $l_{d}$
given by equation (\ref{ld}). The surface density is therefore amplified by a
factor $(R/l_{d})^{2}$ depending on the size of the particles. Using equations
(\ref{lRlight})(\ref{lRheavy}), this amplification can be written explicitely
\begin{equation}
\biggl ({\sigma_{d}^{vort}\over \sigma_{d}}\biggr )_{conc.}={3\over
16\alpha}\sqrt{\gamma+1}{\Omega\over\xi}\quad ({\rm light \ particles})
\label{ampliflight}
\end{equation}
\begin{equation}
\biggl ({\sigma_{d}^{vort}\over \sigma_{d}}\biggr )_{conc.}={3\over
16\alpha}\sqrt{\gamma+1}  \biggl ({\xi\over\Omega}\biggr )^{1/2}\ ({\rm heavy \
particles})
\label{amplifheavy}
\end{equation}
The amplification is maximum for $\xi\sim\Omega$ and takes the value
$A^{conc.}_{max}={3\over 16\alpha}\sqrt{\gamma+1}\simeq 150$. As we have seen in
section \ref{sec_application1}, this corresponds to particles of size $a= 30 cm$
and bulk density $\rho_{d}=2 g/cm^{3}$ in the regions of the Earth and Jupiter.
This enhancement is sufficient to satisfy the instability criterion
(\ref{jeans1}) \footnote{We can argue that when $l_{d}$ is comparable to the
sub-disk thickness $H_{d}$, the instability criterion derived in the case of a
thin disk is not applicable anymore. However, according to Jeans criterion, a
volume element of size $H_{p}$ and mass $M_{p}$ is unstable if the velocity
dispersion of the particles $c_{d}^{2}$ is less than their potential energy
${GM_{d}\over H_{d}}$. Since $c_{d}\sim H_{d}\Omega$ and $M_{d}\sim
\sigma_{d}H_{d}^{2}$, this condition returns the criterion (\ref{jeans1}).}. For
microscopic particles, on the contrary, there is no density enhancement. In that
case,  $l_{d}\sim 10^{3}R$ (in the region of the planets) and the particles
rapidely diffuse away from the vortices (see section \ref{sec_application3}).
Gravitational instability will be possible provided that:
\begin{equation}
{\xi\over\Omega}<{3\over 16\alpha}\sqrt{\gamma+1}{1\over A_{c}}\simeq 1.53
\label{range2}
\end{equation}   
On table \ref{table}, we report, as a function of the heliocentric distance, the
minimum size of the particles which satisfy this criterion (see also figures
\ref{fig_xivsa}  and \ref{fig_ampconcvsa}). These results indicate that
particles must have grown up to some centimeters to trigger the gravitational
instability. Therefore, sticking processes are needed to reach this range of
sizes (recall that this was claimed to be not necessary in the initial
Safronov-Goldreich-Ward scenario). 

\begin{figure}
    \includegraphics{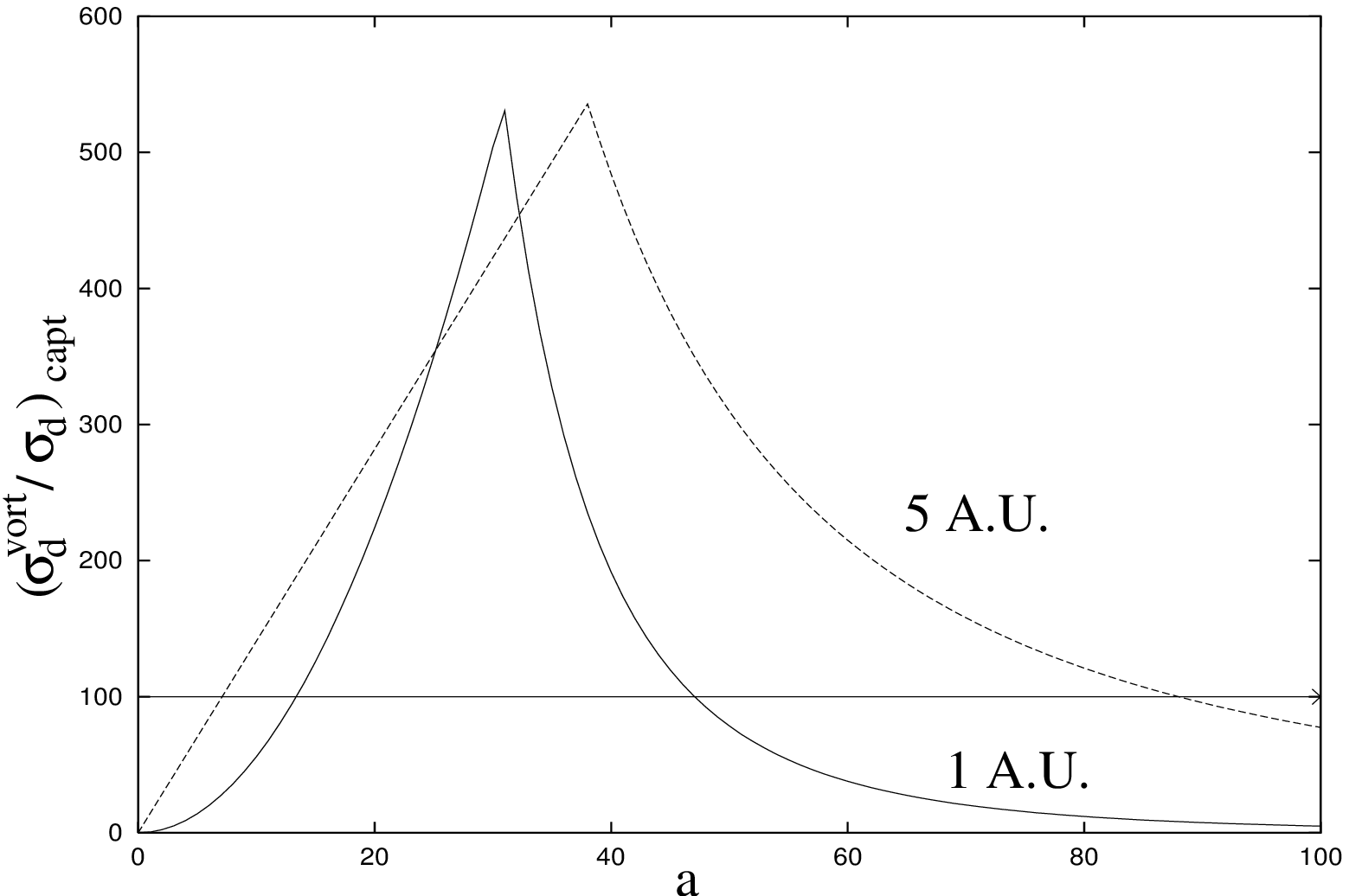}   
\caption[]{Enhancement of the dust surface density inside the vortices (due to
capture) as a function of the size of the particles at $r=1A.U$ and $r=5A.U$
($\rho_{s}=2 g/cm^{3}$).  }
         \label{fig_ampcaptvsa}
   \end{figure}

In conclusion, by allowing a local enhancement of the particle surface density,
the vortices can favour the formation of planetesimals by gravitational
instability. This rehabilitates the Safronov-Goldreich-Ward theory at certain
prefered locations of the disk (i.e inside the vortices) and for sufficiently
large (decimetric) particles.  A sufficient enhancement is achieved simply by
the horizontal {\it concentration} of the dust layer in the vortex (a process
similar to the vertical sedimentation). However, this mechanism alone is not
sufficient to produce enough planetesimals to form the planets. The vortices
must also {\it capture} the surrounding mass. This is another source for density
enhancement (Barge \& Sommeria 1995). Returning to equation (\ref{mlife}), we
find that the average surface density of the particles collected by a vortex
during its lifetime is
\begin{equation}
\biggl ({\langle \sigma_{d}^{vort}\rangle\over \sigma_{d}}\biggr
)_{capt.}={3\over 2\pi}(\Omega t_{life})f^{2}(\xi)
\label{capturedensity}
\end{equation}   
with $\Omega t_{life}\sim 1122$ (see section \ref{sec_application1}). The
maximum amplification, reached by particles with $\xi\sim\Omega$, is
$A^{capt.}_{max}={3\over 2\pi}(\Omega t_{life})\simeq 536$ a little bit larger
than the previous value (recall however that $t_{life}$ is not known precisley).
Gravitational instability will be possible for particles whose friction
parameter satisfies 
\begin{equation}
{\xi\over\Omega}<{3\over 2\pi}(\Omega t_{life}){1\over A_{c}}\simeq 5.4
\label{range3}
\end{equation}   
See also table \ref{table} and figures \ref{fig_xivsa}, \ref{fig_ampcaptvsa} for
the same criterion expressed in terms of the size of the particles.

If we now take into account both the concentration effect and the capture
process, we obtain an amplification
\begin{equation}
{\sigma_{d}^{vort}\over \sigma_{d}}={M_{life}\over \sigma_{d}\pi
l_{d}^{2}}=\biggl ({\sigma_{d}^{vort}\over \sigma_{d}}\biggr )_{conc.}\biggl
({\langle \sigma_{d}^{vort}\rangle\over \sigma_{d}}\biggr )_{capt.}
\label{captconc}
\end{equation}
with a maximum value $A_{max}=A^{conc.}_{max}A^{capt.}_{max}\sim 10^{5}$. The
range of particles which can collapse is enlarged: 
\begin{equation}
{\xi\over\Omega}<\biggl ({9\over 32\pi}{\sqrt{\gamma+1}\over\alpha}(\Omega
t_{life}){1\over A_{c}}\biggr )^{1/2}\simeq 28.6
\label{range4}
\end{equation}
but, even in this optimistic situation, the particles must have reached
relatively large sizes to trigger the gravitational instability (see table
\ref{table}). Of course, if the vortex lifetime is increased, smaller particles
have the possibility to collapse since the vortex captures more mass. In fact,
this is not completely correct because the  previous results assume that the
escape of particles due to turbulent fluctuations can be neglected. This is not
always the case (in particular for small particles) and this problem is now
considered in detail.

\section{The rate of escape}
\label{sec_escape}

\subsection{Formulation of the problem}
\label{sec_problem}

The diffusion equation (\ref{fp}) is similar, in structure, with the
Kramers-Chandrasekhar equation:
\begin{equation}
{\partial f\over\partial t}={\partial \over\partial {\bf v}}\biggl (D{\partial
f\over\partial {\bf v}}+\xi f {\bf v}\biggr )
\label{KC}
\end{equation}     
introduced  in the case of colloidal suspensions and in stellar dynamics
(Kramers 1940, Chandrasekhar 1943a,b). In this equation, $f({\bf v},t)$ governs
the velocity distribution of the particles in the system. The first term in the
r.h.s  is a pure diffusion and the second term is a {\it dynamical friction}.
These terms model the encounters between stars or the collisions between the
colloidal particles and the fluid molecules. Comparing equations (\ref{fp}) and
(\ref{KC}), we see that the position in (\ref{fp}) plays the role of the
velocity in (\ref{KC}) and the capture time $t_{capt}$ the role of the friction
time $\xi^{-1}$. In particular the friction force and the drift term are {\it
linear} in ${\bf v}$ and ${\bf r}$ respectively. Equation (\ref{KC}) was used by
Chandrasekhar (1943b) to study the evaporation of stars in globular clusters
(this is similar to the Kramers problem for the escape of colloidal suspensions
over a potential barrier). Due to collisions, some stars may acquire very high
energies and escape from the system (being ultimately captured by the gravity of
nearby objects). Similarly, in our situation, turbulent fluctuations allow some
dust particles to diffuse towards higher and higher radii and finally leave the
vortex (being eventually transported by the local Keplerian shear). In each
case, the friction force or the drift acts against the diffusion and can reduce
significantly  the escape process.

We try now to evaluate the rate of particles that leave the vortex on account of
turbulent fluctuations. To that purpose, we formulate the problem in terms of
the density probability $W({\bf r},t)=W(|{\bf r}|,t)$ that a particle located
initially in the annulus between $|{\bf r}|=r_{0}$ and $r_{0}+dr_{0}$  will be
found in the surface element around ${\bf r}$ at time $t$. According to equation
(\ref{fp}), the time evolution of the probability $W({\bf r},t)$ is given by
\begin{equation}
{\partial W\over\partial t}=\nabla\biggl (D\nabla W+{W\over t_{capt}}{\bf
r}\biggr )
\label{diffeqW}
\end{equation} 
with initial condition
\begin{equation}
W({\bf r},t)={\delta(|{\bf r}|-{r}_{0})\over 2\pi r_{0}} \quad {\rm as}\quad
t\rightarrow 0
\label{initcond}
\end{equation} 
where $\delta$ stands for Dirac's $\delta$-function. We assume that when the
particle reaches the vortex boundary at $|{\bf r}|=R$, it is immediately
transported away be the Keplerian shear. In other words, we adopt the boundary
condition: 
\begin{equation}
W({\bf r},t)=0 \quad {\rm for}\quad  |{\bf r}|=R \quad {\rm for \ all} \quad t>0
\label{boundary}
\end{equation}
We call 
\begin{equation}
{\bf J}=-\biggl (D\nabla W+{W\over t_{capt}}{\bf r}\biggr )
\label{current}
\end{equation}
the current of probability, i.e ${\bf J} dl \hat{\bf n}$ gives the probability
that a particle crosses an element of length $dl$ between $t$ and $t+dt$
($\hat{\bf n}$ is a unit vector normal to the element of length under
consideration).

We first introduce the probability $p(r_{0},t)dt$ that a particle located
initially in the annulus between $|{\bf r}|=r_{0}$ and $r_{0}+dr_{0}$ reaches
for the first time the vortex boundary between $t$ and $t+dt$. According to what
was just said concerning the interpretation of (\ref{current}), we have:
\begin{equation}
p(r_{0},t)=\oint_{|{\bf r}|=R}{\bf J}\hat{\bf n}dl=-\biggl (2\pi Dr{\partial
W\over\partial r}\biggr )_{r=R}
\label{p}
\end{equation}
The total probability $Q(r_{0},t)$ that the particle has reached the vortex
boundary between $0$ and $t$ is
\begin{equation}
Q(r_{0},t)=\int_{0}^{t}p(r_{0},t')dt'
\label{Q0t}
\end{equation}
Finally, we average $Q(r_{0},t)$ over an appropriate range of initial positions
in order to get the expectation $Q(t)$ that the particle has left the vortex at
time $t$. We have
\begin{equation}
Q(t)=\int Q(r_{0},t)\mu(r_{0})2\pi r_{0}dr_{0}
\label{Q}
\end{equation} 
where $\mu(|{\bf r}_{0}|)$ governs the initial probability distribution of the
particles in the vortex. In terms of the function $Q$, the rate of escape of the
particles can be written
\begin{equation}
{1\over t_{esc}}={1\over (1-Q)}{dQ\over dt}
\label{escape}
\end{equation} 

As mentioned already, this problem is similar to the diffusion of colloidal
suspensions over a potential barrier or to the evaporation of stars in globular
clusters. As will soon become apparent, it reduces to solving a pseudo
Schr\"odinger equation for a quantum oscillator in a ``box''. In this analogy,
the rate of escape appears to be related to the fundamental eigenvalue of the
quantum problem. An explicit expression for the rate of escape can be obtained
in two limits: when $\xi\rightarrow 0$ or $\xi\rightarrow\infty$, the drift term
can be ignored and the Fokker-Planck equation reduces to a pure diffusion
equation (section \ref{sec_ignored}). On the other hand, when $\xi\simeq
\Omega$, the drift term is dominant and a perturbation approach inspired by the
work of Sommerfeld and Chandrasekhar can be implemented to determine the ground
state of the artificially limited quantum oscillator (sections
\ref{sec_schrodinger} and \ref{sec_ground}).

\subsection{The rate of escape when the drift term is ignored}
\label{sec_ignored}

When the drift term can be ignored (i.e for very light or very heavy particles,
see inequalities (\ref{range1})(\ref{range1bis})), the Fokker-Planck equation
(\ref{diffeqW}) reduces to a pure diffusion equation 
\begin{equation}
{\partial W\over\partial t}=D\Delta W
\label{diffW}
\end{equation} 
which has to be solved in a circular domain with boundary conditions
(\ref{initcond}) and (\ref{boundary}). The solution of this classical problem is
\begin{eqnarray}
W={1\over \pi R^{2}}\sum_{n=1}^{+\infty}{1\over
J_{1}^{2}(\alpha_{0n})}J_{0}\biggl (\alpha_{0n}{r_{0}\over R}\biggr )
\nonumber\\ 
\times J_{0}\biggl (\alpha_{0n}{r\over R}\biggr )
e^{-{D\alpha_{0n}^{2}\over R^{2}}t}
\label{Wdiffexp}
\end{eqnarray} 
where $J_{m}$ is Bessel function of order $m$ and the $\alpha_{0n}$'s denote the
roots of Bessel function $J_{0}$. The probability $p(r_{0},t)$ that a particle
with an initial position $r_{0}$ has reached the vortex boundary between $t$ and
$t+dt$ is 
\begin{equation}
p(r_{0},t)={2D\over R^{2}}\sum_{n=1}^{+\infty}{\alpha_{0n}\over
J_{1}(\alpha_{0n})}J_{0}\biggl (\alpha_{0n}{r_{0}\over R}\biggr
)e^{-{D\alpha_{0n}^{2}\over R^{2}}t}
\label{pexpdiff}
\end{equation} 
and the total probability that the particle has escaped during the interval
$(0,t)$ is
\begin{eqnarray}
Q(r_{0},t)={2}\sum_{n=1}^{+\infty}{1\over \alpha_{0n}
J_{1}(\alpha_{0n})}J_{0}\biggl (\alpha_{0n}{r_{0}\over R}\biggr )\nonumber\\
\times \biggl  (1-e^{-{D\alpha_{0n}^{2}\over R^{2}}t}\biggr  )
\label{Q0texp}
\end{eqnarray} 
Averaging the foregoing expression over all $r_{0}$'s in the range $\lbrack
0,R\rbrack$ with equal probability $\mu({\bf r}_{0})=1/ \pi R^{2}$ (this
corresponds to an initially homogeneous distribution of particles in the
vortex), we obtain
\begin{equation}
Q(t)=4\sum_{n=1}^{+\infty}{1\over \alpha_{0n}^{2}}\biggl
(1-e^{-{D\alpha_{0n}^{2}\over R^{2}}t}\biggr )
 \label{Qexp}
\end{equation} 
To sufficient accuracy, we can keep only the first term in the series. The
expectation that the particle has left the vortex at time $t$ is therefore:
\begin{equation}
Q(t)\simeq {4\over \alpha_{01}^{2}}\biggl (1-e^{-{D\alpha_{01}^{2}\over
R^{2}}t}\biggr )
\label{Qexpfin}
\end{equation} 
Since $\alpha_{01}\simeq 2.40482...$, this term represents $\sim 70\%$ of the
value of the series (\ref{Qexp}) and the approximation (\ref{Qexpfin}) is
reasonable. 

In conclusion, when the drift term is ignored, we find that the escape time is
\begin{equation}
t_{esc}={R^{2}\over D\alpha_{01}^{2}}
\label{tesc}
\end{equation}
It corresponds, typically, to the time needed by the particles to diffuse over a
distance $\sim R$, the vortex size. For light particles, using (\ref{Dlight}),
we obtain explicitely
\begin{equation}
\Omega t_{esc}={\sqrt{\gamma+1}\over \alpha_{01}^{2}\alpha}\quad ({\rm light \
particles})
\label{tesclight}
\end{equation} 
and for heavy particles
\begin{equation}
\Omega t_{esc}={\sqrt{\gamma+1}\over \alpha_{01}^{2}\alpha}\biggl
({\Omega\over\xi}\biggr )^{1/2}\quad ({\rm heavy \ particles})
\label{tescheavy}
\end{equation} 
We shall come back to these expressions in section \ref{sec_application3}.

\subsection{The effective Schr\"odinger equation}
\label{sec_schrodinger}

We now return to the general problem for the rate of escape when proper
allowance is made for the drift. We find it convenient to introduce the
notations
\begin{equation}
\tau={t\over t_{capt}}\quad {\rm and} \quad {\mb \rho}={1\over\sqrt{2 D t_{capt}
}}{\bf r}
\label{notation1}
\end{equation} 
or, in words, the time is measured in terms of the capture time and the
distances are normalized by the concentration length (\ref{ld}). We let also:
\begin{equation}
w=2 D t_{capt} W
\label{notation2}
\end{equation} 
and
\begin{equation}
\rho_{\infty}={1\over\sqrt{2 D t_{capt}}} R
\label{notation3}
\end{equation} 
In terms of these new variables, the problem (\ref{diffeqW})(\ref{initcond}) and
(\ref{boundary}) takes the form:
\begin{equation}
{\partial w\over\partial \tau}={1\over 2}\Delta w+\nabla (w{\mb\rho})
\label{diffeqWaxn}
\end{equation}  
\begin{equation}
w({\mb \rho},\tau)={\delta({\rho}-{\rho}_{0})\over 2\pi \rho_{0}} \quad {\rm
as}\quad  \tau\rightarrow 0
\label{initcondaxn}
\end{equation} 
\begin{equation}
w({\mb \rho},\tau)=0 \quad {\rm for} \quad \rho=\rho_{\infty} \quad {\rm for \
all}\quad  \tau>0
\label{boundaryaxn}
\end{equation}
With the change of variables
\begin{equation}
w=\psi e^{-\rho^{2}/2}
\label{wpsi}
\end{equation}
we can transform the Fokker-Planck equation (\ref{diffeqWaxn}) into a
Schr\"odinger equation (with imaginary time) for a quantum oscillator:
\begin{equation}
{\partial \psi\over\partial \tau}={1\over 2}\Delta \psi+(1-{1\over
2}\rho^{2})\psi
\label{schrodinger}
\end{equation}
However, contrary to the standard quantum problem, equation (\ref{schrodinger})
has to be solved  in a bounded domain of size $\rho_{\infty}$ with the boundary
condition (\ref{boundaryaxn}). In other words, our problem consists in
determining the characteristic functions of a quantum oscillator in a ``box''.

First, we notice that a separation of the variables can be effected by the
substitution
\begin{equation}
\psi=\phi(\rho)e^{-\lambda\tau}
\label{separable}
\end{equation}
where $\lambda$ is, for the moment, an unspecified constant. This transformation
reduces the Schr\"odinger equation in a second order ordinary differential
equation:
\begin{equation}
{d^{2}\phi\over d\rho^{2}}+{1\over\rho}{d\phi\over
d\rho}+(2+2\lambda-\rho^{2})\phi=0
\label{phi}
\end{equation}
Let $\lbrace\phi_{n}\rbrace$ be the solutions of this differential equation
satisfying the boundary condition $\phi_{n}(\rho_{\infty})=0$ and $\lambda_{n}$
the corresponding eigenvalues. The eigenfunctions form a complete set of
orthogonal functions for the scalar product
\begin{equation}
\langle f g\rangle=\int_{0}^{\rho_{\infty}} f(\rho)g(\rho) 2\pi\rho d\rho
\label{scalar}
\end{equation}
This system can be further normalized, i.e $\langle \phi_{n}\phi_{m}\rangle
=\delta_{nm}$. Any function $f(\rho)$ satisfying the boundary condition
(\ref{boundaryaxn}) can be expanded on this basis, and we have the formula:
\begin{equation}
f(\rho)=\sum_{n}\langle f\phi_{n}\rangle \phi_{n}
\label{expansion}
\end{equation}
In particular:
\begin{equation}
\delta(\rho-\rho_{0})=\sum_{n}2\pi\rho_{0}\phi_{n}(\rho_{0})\phi_{n}(\rho)
\label{delta}
\end{equation}

The general solution of the problem (\ref{diffeqWaxn}) (\ref{boundaryaxn}) can
be expressed in the form 
\begin{equation}
w(\rho,\tau)=\sum_{n}A_{n}e^{-\lambda_{n}\tau}e^{-\rho^{2}/2}\phi_{n}(\rho)
\label{wrt}
\end{equation}
where the coefficients $A_{n}$ are determined by the intial condition
(\ref{initcondaxn}), using the expansion (\ref{delta}) for the
$\delta$-function. We obtain:
\begin{equation}
w(\rho,\tau)=e^{-(\rho^{2}-\rho_{0}^{2})/2}\sum_{n}
e^{-\lambda_{n}\tau}\phi_{n}(\rho)\phi_{n}(\rho_{0})
\label{wrtexp}
\end{equation}
Using the foregoing solution for $w$, the probability that a particle initially
located at $\rho_{0}$ leaves the vortex between $\tau$ and $\tau+d\tau$ is [see
Eq. (\ref{p})]:
\begin{eqnarray}
p(\rho_{0},\tau)=-\pi\rho_{\infty}e^{-(\rho_{\infty}^{2}-\rho_{0}^{2})/2}
\nonumber\\
\times \sum_{n}e^{-\lambda_{n}\tau}{d\phi_{n}\over
d\rho}(\rho_{\infty})\phi_{n}(\rho_{0})
\label{pexp}
\end{eqnarray}
The probability $Q(\rho_{0},\tau)$ that the particle has left the vortex at time
$\tau$ is therefore [see Eq. (\ref{Q0t})]:
\begin{eqnarray}
Q(\rho_{0},\tau)=-\pi\rho_{\infty}e^{-(\rho_{\infty}^{2}-\rho_{0}^{2})/2}
\nonumber\\
\times\sum_{n}{1-e^{-\lambda_{n}\tau}\over\lambda_{n}}{d\phi_{n}\over
d\rho}(\rho_{\infty})\phi_{n}(\rho_{0})
\label{Qdexp}
\end{eqnarray}
Finally, to obtain $Q(\tau)$,  we have to take the average of the foregoing
expression over the relevant range of $\rho_{0}$. To make the solution explicit,
it remains to determine the eigenfunctions $\phi_{n}$ and eigenvalues
$\lambda_{n}$ of the quantum oscillator.

\subsection{The ground state of the quantum oscillator}
\label{sec_ground}

The dependance of the eigenvalues $\lambda_{n}$ on the size of the ``box'' can
be obtained from a procedure developed by Sommerfeld in his studies of the
Kepler problem and the problem of the rotator in the quantum theory with
``artificial'' boundary conditions (Sommerfeld \& Welker 1938, Sommerfeld \&
Hartmann 1940). This was used later on by Chandrasekhar (1943) to determine the
rate of escape of stars from globular clusters, from which our study is
inspired.

\begin{figure}
    \includegraphics{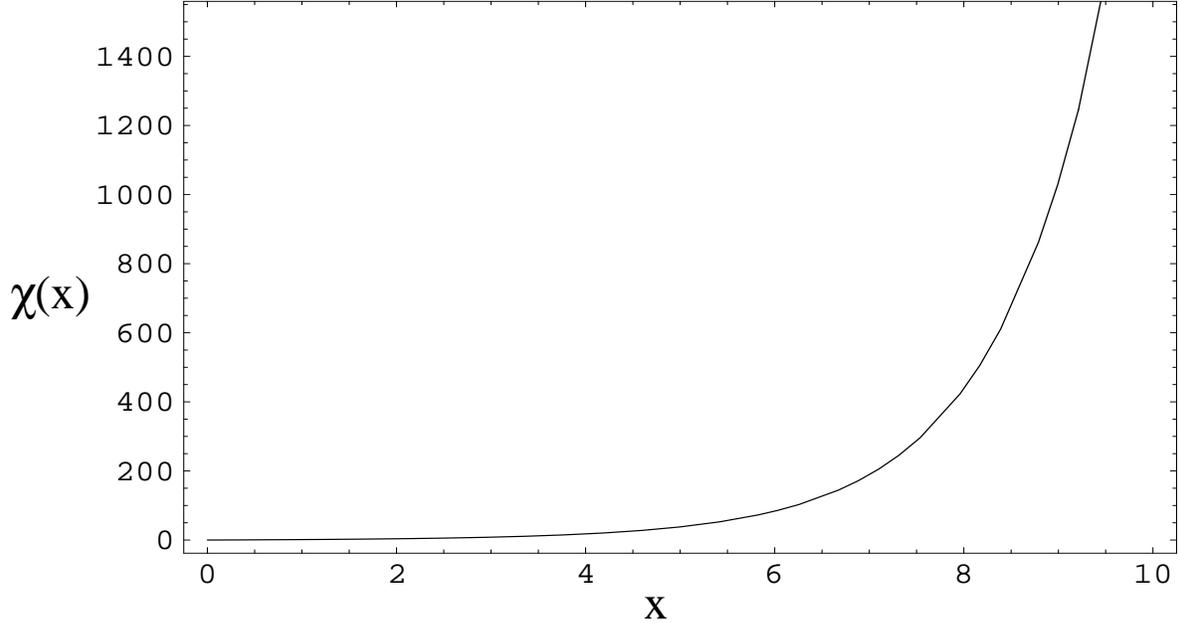}  
      \caption[]{The function $\chi(x)$. }
         \label{fig_chivsx}
   \end{figure}

With the change of variables 
\begin{equation}
x=\rho^{2}\quad 
\label{xrho2}
\end{equation}
equation (\ref{phi}) becomes:
\begin{equation}
x{d^{2}\phi\over dx^{2}}+{d\phi\over dx}+\biggl ( {1\over 2}+{\lambda\over
2}-{x\over 4}\biggr )\phi=0
\label{xphi}
\end{equation}
Another change of variables:
\begin{equation}
\phi=f e^{-x/2} 
\label{phif}
\end{equation}
transforms equation (\ref{xphi}) into  Kummer's equation
\begin{equation}
x{d^{2}f\over dx^{2}}+(1-x){df\over dx}+{\lambda\over 2} f=0
\label{fkum}
\end{equation}
In the case of a ``free'' oscillator ($\rho_{\infty}\rightarrow\infty$), it is
well-known that the eigenvalues are $\lambda_{n}=2n$ (with $n=0,1,...$) and the
eigenfunctions are proportional to Laguerre polynomials $f_{n}=A L_{n}^{(0)}$.
In a bounded domain ($\rho_{\infty}<\infty$), the eigenvalues will be different
but if $\rho_{\infty}$ is sufficiently large, we expect that the difference will
be small. Therefore, we expect $\lambda_{n}\simeq 2n$. Accordingly,  for values
of $\tau$ of the order of unity, or greater, the first term in the series
(\ref{Qdexp}) will provide ample accuracy. Thus
\begin{eqnarray}
Q(\rho_{0},\tau)\simeq
-\pi\rho_{\infty}e^{-(\rho_{\infty}^{2}-\rho_{0}^{2})/2}\nonumber\\
\times{1\over \lambda_{0}}({1-e^{-\lambda_{0}\tau}}){d\phi_{0}\over
d\rho}(\rho_{\infty})\phi_{0}(\rho_{0})
\label{Qexpapp}
\end{eqnarray}  
We have therefore to determine the {\it ground state} of our artificially
limited quantum oscillator. To that purpose, we expand $f$ in a series:
\begin{equation}
f=\sum_{n}a_{n}x^{n}
\label{fseries}
\end{equation} 
and substitute this expansion into equation (\ref{fkum}). Identifying term by
term, we obtain the recursion formula:
\begin{equation}
a_{n+1}={n-{\lambda\over 2}\over (n+1)^{2}}a_{n}
\label{recursion}
\end{equation}  
which is easily reduced to
\begin{equation}
a_{n}={ (n-1-{\lambda\over 2})(n-2-{\lambda\over 2})...(-{\lambda\over 2})\over
(n!)^{2}}a_{0}
\label{an}
\end{equation}  
We know already that the eigenvalue $\lambda_{0}$ of our artificial quantum
oscillator is very small. Keeping only terms of order $\lambda_{0}$ in the
products (\ref{an}), we obtain:
\begin{eqnarray}
a_{n}\simeq {(n-1) (n-2)...1 (-{\lambda_{0}\over 2})\over
(n!)^{2}}a_{0}\nonumber\\ 
=-{(n-1)!\over (n!)^{2}}{\lambda_{0}\over 2}
a_{0}=-{a_{0}\over n!n}{\lambda_{0}\over 2}
\label{anapp}
\end{eqnarray}
The ground state function can therefore be written
\begin{equation}
f_{0}(x)=a_{0} (1-{\lambda_{0}\over 2}\chi(x) )
\label{ffinal}
\end{equation} 
with
\begin{equation}
\chi(x)=\sum_{n=1}^{+\infty} {x^{n}\over n!n}=E_{i}(x)-\ln x-\gamma
\label{chi}
\end{equation}
where
\begin{equation}
E_{i}(x)=P\int_{-\infty}^{x} {e^{t}\over t}dt
\label{Ei}
\end{equation}
is the Exponential integral and  $\gamma=0.577215...$  the Euler constant. The
function $\chi(x)$ is plotted on figure \ref{fig_chivsx}. The fundamental
eigenvalue $\lambda_{0}$ is determined by the condition
$f_{0}(\rho_{\infty}^{2})=0$, i.e
\begin{equation}
\lambda_{0}={2\over\chi(\rho_{\infty}^{2})}
\label{epsilon}
\end{equation}
Returning to the original eigenfunction $\phi_{0}(\rho)$, we have established
that
\begin{equation}
\phi_{0}(\rho)=a_{0}(1-{\lambda_{0}\over 2}\chi(\rho^{2}))e^{-\rho^{2}/2}
\label{phi0}
\end{equation}
where $a_{0}$ is a normalizing factor. The final result can therefore be
expressed in the form:
\begin{equation}
Q(\tau)=A (1-e^{-\lambda_{0}\tau})
\label{Qdfinal}
\end{equation}
with
\begin{equation}
A=-{\pi\rho_{\infty}\over \lambda_{0}}e^{-\rho_{\infty}^{2}/2}{d\phi_{0}\over
d\rho}(\rho_{\infty})\overline{e^{\rho_{0}^{2}/2}\phi_{0}(\rho_{0})}
\label{A}
\end{equation}
In the expression for the amplitude, the bar indicates an average over the
relevant range of $\rho_{0}$. In practice, $A$ is close to unity but its precise
value determines to which accuracy the approximation consisting in keeping only
the dominant term in the series (\ref{Qdexp}) is justified. For sufficiently
large $\rho_{\infty}$'s, this will always be the case, so we shall take
\begin{equation}
Q(\tau)\simeq 1-e^{-\lambda_{0}\tau}
\label{Qdfinalapp}
\end{equation}
This expression is consistent with the physical condition $Q(\tau)\rightarrow 1$
as $\tau\rightarrow \infty$.

\subsection{Application to the solar nebula}
\label{sec_application3}

\begin{figure}
    \includegraphics{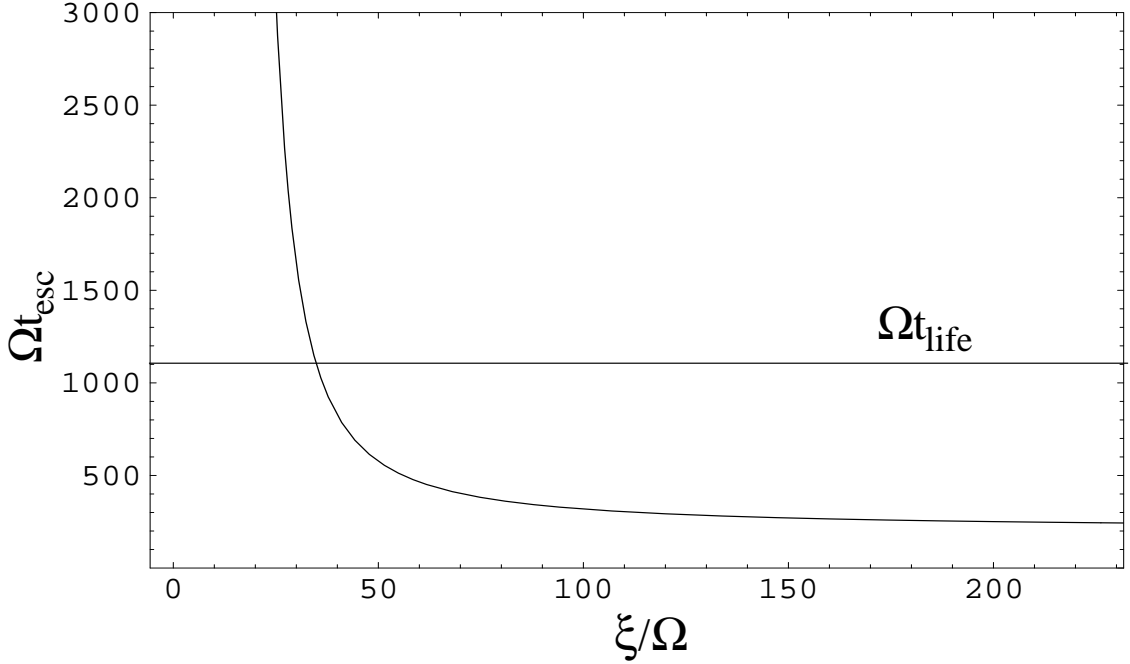} 
\caption[]{Escape time of the particles as a function of their friction
parameter. }
         \label{fig_tescvsxilight}
   \end{figure}

\begin{figure}
    \includegraphics{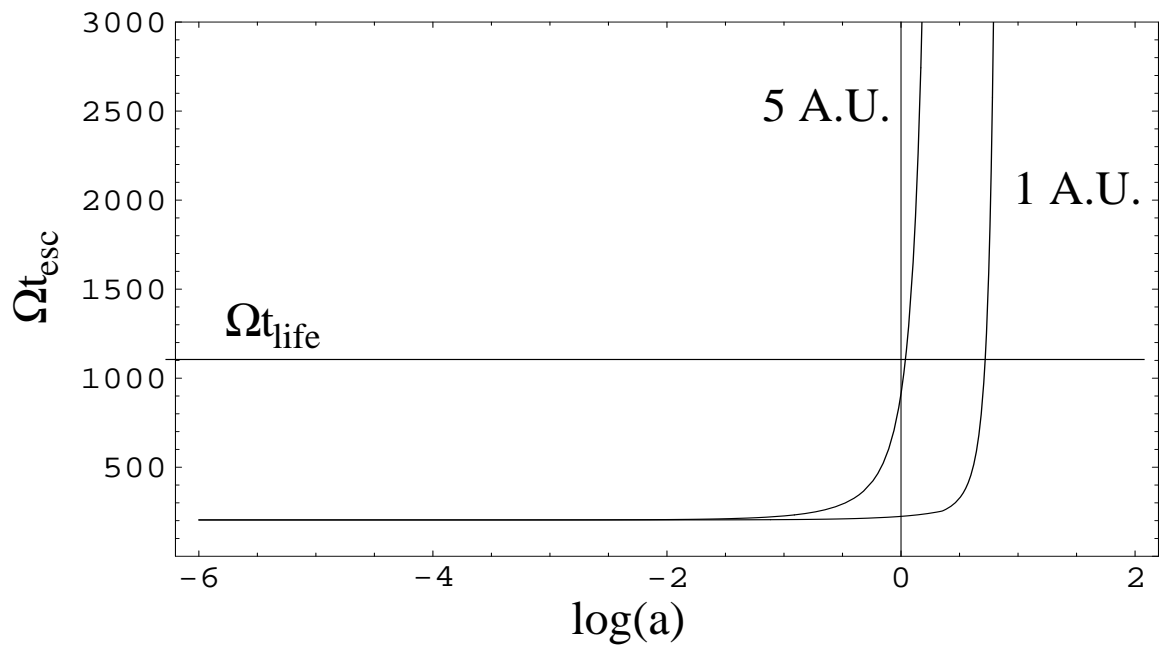}   
\caption[]{Escape time of the particles as a function of their size at $r=1A.U$
and $r=5A.U$ ($\rho_{s}=2g/cm^{3}$). }
         \label{fig_tescvsa}
   \end{figure}

Let $M_{0}$ denote the total mass of particles present in the vortex at time
$t=0$. If we assume no renewal from the outside then, on account of turbulent
fluctuations, some particles will escape the vortex and, according to equation
(\ref{Qdfinalapp}), the mass will decay like:
\begin{equation}
M(t)=M_{0}e^{-\lambda_{0} t/t_{capt}}
\label{Mt}
\end{equation}
The ``evaporation'' will take place on a typical time
\begin{equation}
t_{esc}={1\over \lambda_{0}} t_{capt}
\label{tevapd}
\end{equation}
where $\lambda_{0}$ is given by equation (\ref{epsilon}). Note that both
$\lambda_{0}$ and $t_{capt}$ depend on the friction parameter $\xi$ of the
particles. For very light or very heavy particles, $\rho_{\infty}\rightarrow 0$
and $\lambda_{0}\simeq 2/\rho_{\infty}^{2}$. With the definition
(\ref{notation3}) we find:
\begin{equation}
t_{esc}={R^{2}\over 4 D}
\label{tevapdexp}
\end{equation}
in good agreement with the result (\ref{tesc}) obtained when the drift term is
ignored. This shows that equation (\ref{tevapd}) can be used with good accuracy
for all types of particles. Substituting equations
(\ref{lRlight})(\ref{lRheavy}) and (\ref{taulight})(\ref{tauheavy}) into
equation (\ref{tevapd}), we obtain explicitely:
\begin{equation}
\Omega t_{esc}={4\over 3}{\xi\over\Omega}\chi\biggl ({3\sqrt{\gamma+1}\over
16\alpha}{\Omega\over\xi}\biggr )\quad ({\rm light \ particles})
\label{esclight}
\end{equation} 
\begin{equation}
\Omega t_{esc}={4\over 3}{\Omega\over\xi}\chi\biggl ({3\sqrt{\gamma+1}\over
16\alpha}\biggl ({\xi\over\Omega}\biggr )^{1/2}\biggr )\ ({\rm heavy \
particles})
\label{escheavy}
\end{equation} 
where $\chi$ is the function defined by equation (\ref{chi}). When
$\xi\rightarrow \infty$, equation (\ref{esclight}) tends to
\begin{equation}
\Omega t_{esc}={\sqrt{\gamma+1}\over 4\alpha}\simeq 200
\label{xiinfty}
\end{equation} 
Very light particles are not concentrated in the vortices and they diffuse away
after only $\sim 30$ rotation periods. This is the {\it minimum} escape time. By
contrast, for optimal particles with $\xi\sim\Omega$, the concentration reduces
considerably the evaporation and we find 
\begin{equation}
\Omega t_{esc}={4\over 3}{\xi\over\Omega}\chi\biggl ({3\sqrt{\gamma+1}\over
16\alpha}\biggr )\sim 10^{64}
\label{xiopt}
\end{equation} 
The particles are so much concentrated in the vortices ($l_{d}\sim 0.1 R$) that
the turbulent fluctuations are not sufficient to allow them to reach the edge of
the vortex. Therefore, their escape is completely negligible. Finally, for
$\xi\rightarrow 0$, we have
\begin{equation}
\Omega t_{esc}={\sqrt{\gamma+1}\over 4\alpha}\biggl ({\Omega\over \xi}\biggr
)^{1/2}
\label{xizero}
\end{equation} 
For very heavy particles, the escape time goes to infinity because the particles
are not coupled to the gas and therefore not affected by diffusion. Recall,
however, that our study is not well suited for very heavy particles. As
discussed in section \ref{sec_capturetime}, these particles are more likely to
cross the vortices without being captured. The curve $\Omega t_{esc}$ versus
$\xi/\Omega$ is potted on figure \ref{fig_tescvsxilight} (see also figure
\ref{fig_tescvsa} for the dependance of $\Omega t_{esc}$ on the size of the
particles). The asymptotic regimes  (\ref{xiinfty}) and (\ref{xizero}) are
obtained for particles with $\xi/\Omega>153$ and $\xi/\Omega<4\ 10^{-5}$
respectively. For these particles, $l_{d}> R$ (see inequalities (\ref{range1})
and (\ref{range1bis})), so there is no concentration. Therefore, the asymptotic
limits (\ref{xiinfty}) and (\ref{xizero}) agree with the results
(\ref{tesclight}) and  (\ref{tescheavy}) obtained when the drift term is
ignored.

These results assume that there is no renewal of the particles in the vortices.
If we now take into account a capture process like in section
\ref{sec_capturerate}, we are led to consider the balance equation:
\begin{equation}
{dM\over dt}={3\over 2}\sigma_{d}\Omega R^{2}f^{2}(\xi)-{1\over t_{esc}}M
\label{balance}
\end{equation} 
The first term in the r.h.s accounts for a flux of particles inside the vortex
(see equation (\ref{mrate})) and the second term for an exponential decay of the
particle number due to ``evaporation'' (see equation (\ref{Mt})).

For particles which can experience gravitational collapse (see inequalities
(\ref{range2})(\ref{range3}) and (\ref{range4})), the escape time is much longer
than the vortex lifetime (see figure \ref{fig_tescvsxilight}). In that case,
evaporation can be neglected and equation (\ref{balance}) reduces to equation
(\ref{mrate}). The maximum mass captured by the vortex is determined by its
lifetime and the results of sections \ref{sec_application1} and
\ref{sec_application2} are unchanged.

\begin{figure}
      \includegraphics{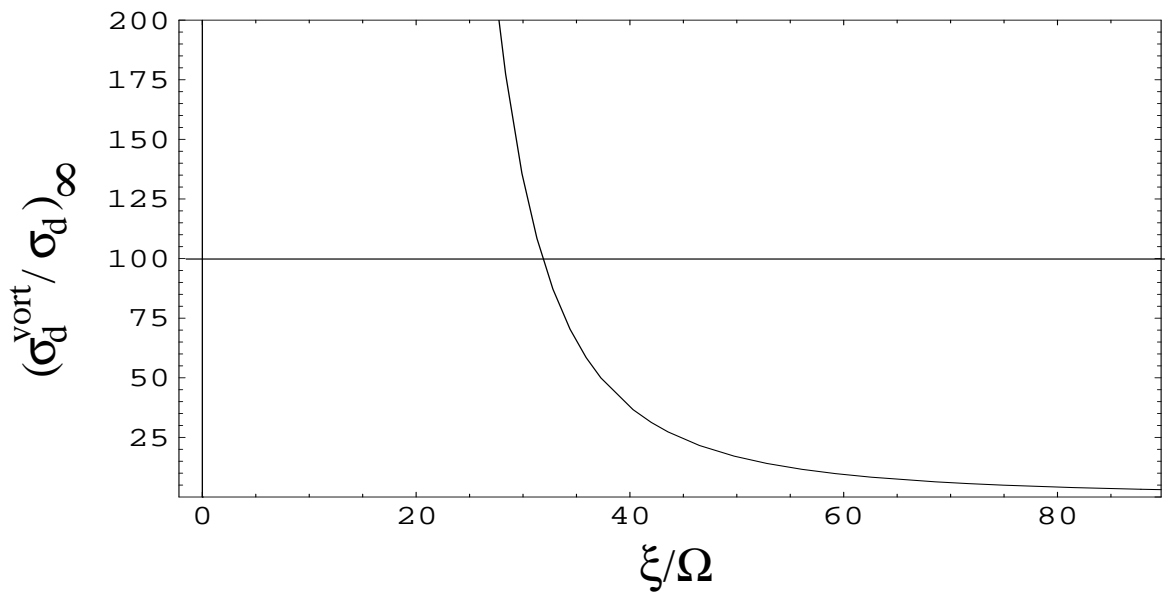} 
\caption[]{Enhancement of the dust surface density in the vortices as a function
of the friction parameter when evaporation due to small scale turbulence is
taken into account ($t_{life}\rightarrow \infty$).}
         \label{fig_ampescvsxi}
   \end{figure}

Consider, however, the idealistic situation when $t_{life}\rightarrow\infty$
(recall that we don't know precisely the value of $t_{life}$). For sufficienlty
large times ($t>t_{esc}$), the vortex  will achieve a stationary distribution of
dust particles obtained by setting $dM/dt=0$ in equation (\ref{balance}). This
gives a maximum mass 
\begin{equation}
M_{max}={3\over 2}\sigma_{d}(\Omega t_{esc}) R^{2}f^{2}(\xi)
\label{Mmaxevap}
\end{equation} 
which is now limited by the evaporation time (compare equation (\ref{Mmaxevap})
with equation (\ref{mlife})). The density enhancement in the vortices (taking
into account the concentration effect) is 
\begin{equation}
\biggl ({ \sigma_{d}^{vort}\over \sigma_{d}} \biggr )_{\infty}={3\over
2\pi}(\Omega t_{esc})f^{2}(\xi)\biggl ({R\over l_{d}}\biggr )^{2}
\label{surfdensity}
\end{equation} 
as represented on figure \ref{fig_ampescvsxi}. We find that even if the vortex
had an infinite lifetime, particles with friction parameter $\xi/\Omega>30$
cannot trigger the gravitational instability. This result  implies  (see figure
\ref{fig_xivsa}) that subcentimetric particles cannot form planetesimals even in
the most optimistic case. Sticking processes are necessary to produce larger
particles.

\section{Conclusion}
\label{sec_conclusion}

This article follows the works of Barge \& Sommeria (1995), Tanga {\it et al.}
(1996), Bracco {\it et al.} (1999) and Godon \& Livio (1999c) concerning the
interaction between  dust particles and large-scale vortices in the solar
nebula. The  originality of our study was to start from an exact and realistic
solution of the fluid equations and to provide analytical results  and relevant
parameters for the trapping process. Moreover, we have  considered for the first
time the effect of small-scale turbulent fluctuations on  the motion of the
particles and determined an explicit expression for the escape time by solving a
problem of quantum mechanics. These theoretical results  have been formulated in
the context of Keplerian disks and planet formation  but they can clearly have
applications in other fields of astrophysics  and geophysics, for example the
transport of polluants in the Earth's  atmosphere.

The vortex scenario provides an attractive mechanism to form planetesimals  on
very short time scales. This does not contradict the overall picture of  planet
formation which has been developed over the last decades. On the contrary, it
fills the gap between two domains which were difficult to connect: {\it sticking
processes} are still necessary to produce centimetric particles and {\it
collision  between planetesimals} is of course the main engine of planet's
growth. In  between, the vortex scenario should come at work to facilitate, at
some prefered locations of the disk, the Safronov-Goldreich-Ward instability
which was proposed initially for forming planetesimals.    

The predictions of the vortex model are remarkably consistent with the structure
of the solar system. The capture process is optimal at two prefered locations of
the nebula which correspond, for relevant sizes of the particles (i.e some
centimeters), to the position of telluric and giant planets. The transition
between the two groups of planets happens to coincide with the passage from the
Stokes to the Epstein regime where the gas drag law changes. The asymmetry
between the two optima may be ascribed to the size of the vortices which are
bigger in the outer zone.

Of course, the results discussed here rely upon the existence of vortices in the
disk. Their presence in the solar nebula is reasonable due to the ubiquitous
appearance of vortices in rotating flows and two-dimensional turbulence.
However, numerical simulations and even laboratory experiments are necessary to
ascertain their existence in Keplerian flows. A first step was undertaken by
Bracco {\it et al.} (1998,1999) and Godon \& Livio (1999a,b,c) who observed the
formation of anticyclones developing from  an energetic random initial vorticity
field in a Keplerian flow. More work remains to be done to understand the
generation (and maintenance) of vorticity in such disks (convection, baroclinic
instability...) and to determine the effect of three-dimensionality (Hodgson \&
Brandenburg 1998) on the stability of theses vortices.

\section{Acknowledgments:}

\bigskip I would like to thank P. Barge and J. Sommeria for introducing me to
this fascinating subject and for several illuminating comments. I acknowledge
also very stimulating discussions with A. Provenzale, A. Bracco, P. Tanga and E.
Spiegel. This work was initiated during my postdoctoral stay at the Istituto di
Cosmogeofisica.

\appendix
\section{Elliptic vortex in a uniform shear}
\label{sec_exactsolution}

In his monograph, Saffman (1992) reports the existence of an exact solution of
the incompressible two-dimensional Euler equation consisting of an elliptic
patch of uniform vorticity $\omega$ embedded in a simple shear $\kappa$. Since
this vortex solution is the starting point of our study, we establish in this
appendix the condition (\ref{relation}) for its existence and construct the
corresponding streamfunction. These results will be useful for subsequent
studies.

First, we introduce a cartesian system of coordinates and write the shear under
the form
\begin{equation}
u_{x}=\kappa y
\label{uxs}
\end{equation}
\begin{equation}
u_{y}=0
\label{uys}
\end{equation}
where $\kappa$ is a constant ($\kappa={3\over 2}\Omega$ for the Keplerian shear
considered in section \ref{sec_vortexmodel}). The associated vorticity is
$\omega_{ext}=-\kappa$.

Inside the vortex, the velocity field can be written
\begin{equation}
u_{x}=-{q^{2}\over 1+q^{2}}\omega y
\label{veloellipsex}
\end{equation}
\begin{equation}
u_{y}={1\over 1+q^{2}}\omega x
\label{veloellipsey}
\end{equation}
where $q=a/b$ is the aspect ratio of the elliptic patch ($a$ and $b$ denote the
semi-axis in the $x$ and $y$ directions respectively) and $\omega$ the
vorticity. We can check that the fluid particles move at constant angular
velocity ${q\over 1+q^{2}}\omega$ along concentric ellipses with aspect ratio
$q$. 

For an incompressible two-dimensional flow, it is convenient to introduce a
streamfunction $\psi$ defined by ${\bf u}=-\hat {\bf z}\wedge\nabla\psi$ (where
$\hat {\bf z}$ is a unit vector normal to the flow). For a given vorticity
field, the stramfunction is solution of the Poisson equation
\begin{equation}
\Delta\psi=-\omega
\label{poisson}
\end{equation}
Inside the vortex, we find that
\begin{equation}
\psi=-{1\over 2}{\omega\over 1+q^{2}}(x^{2}+q^{2}y^{2})
\label{psiin}
\end{equation}
where we have taken, by convention,  $\psi=0$ at the centrer of the vortex. 
On the vortex boundary, the streamfunction is constant with value:
\begin{equation}
\psi_{0}=-{1\over 2}{\omega\over 1+q^{2}}a^{2}
\label{psiinbound}
\end{equation}
Outside the vortex, we have to solve the Poisson equation
\begin{equation}
\Delta\psi=\kappa
\label{poissonkappa}
\end{equation}
with boundary conditions at infinity
\begin{equation}
{\partial\psi\over\partial x}\rightarrow 0, {\partial\psi\over\partial
y}\rightarrow \kappa y \quad {\rm for}\quad x,y\rightarrow\infty
\label{bouninfty}
\end{equation}
These boundary conditions insure a continuous matching with the shear
(\ref{uxs})(\ref{uys}) at large distances. Introducing the decomposition 
\begin{equation}
\psi=\phi+{1\over 2}\kappa y^{2}
\label{psiphi}
\end{equation} 
the problem (\ref{poissonkappa})(\ref{bouninfty}) is equivalent  to solving the
Laplace equation 
\begin{equation}
\Delta\phi=0
\label{laplace}
\end{equation}
with boundary conditions at infinity
\begin{equation}
{\partial\phi\over\partial x}\rightarrow 0, {\partial\phi\over\partial
y}\rightarrow 0 \quad {\rm for}\quad x,y\rightarrow\infty
\label{bouninftyphi}
\end{equation}
At this stage, we find it convenient to use elliptic coordinates
\begin{equation}
x=c\cosh\xi\cos\eta
\label{ex}
\end{equation}
\begin{equation}
y=c\sinh\xi\sin\eta
\label{ey}
\end{equation}
with $0\le \xi<\infty$ and $0\le\eta<2\pi$. The vortex boundary is an ellipse
with parameter $\xi_{0}$ satisfying
\begin{equation}
{x^{2}\over c^{2}\cosh^{2}\xi_{0}}+{y^{2}\over c^{2}\sinh^{2}\xi_{0}}=1
\label{xinot}
\end{equation}
This relation determines the semi-axis $a$ and $b$ of the ellipse in terms of
$\xi_{0}$ and $c$: 
\begin{equation}
a=c\cosh\xi_{0}\qquad b=c\sinh\xi_{0} 
\label{ab}
\end{equation}
Alternatively, we have
\begin{equation}
\tanh\xi_{0}={b\over a}={1\over q}\qquad c^{2}=a^{2}-b^{2} 
\label{xi0}
\end{equation}
In terms of elliptic coordinates the Laplace equation (\ref{laplace}) has the
simple form
\begin{equation}
{\partial^{2}\phi\over\partial\xi^{2}}+{\partial^{2}\phi\over\partial\eta^{2}}=0
\label{Le}
\end{equation}
This equation, with the boundary condition (\ref{bouninftyphi}), is solved
easily and we obtain outside the vortex 
\begin{equation}
\psi={1\over 2}\kappa
y^{2}+B\xi+\sum_{n=0}^{+\infty}A_{n}e^{-n\xi}\cos(n\eta+\gamma_{n})
\label{psiext}
\end{equation}
where $B$, $A_{n}$ and $\gamma_{n}$ are some constants. At large distances,
$\xi\rightarrow \ln r$ and $\eta\rightarrow\theta$ (where $r$ and $\theta$ are
polar coordinates) and we recover the usual multipolar expansion of the
streamfunction. The condition that the vortex boundary is a streamline destroys
most of the terms in the series (\ref{psiext}). There remains only
\begin{eqnarray}
\psi={\kappa c^{2}\over 4}\sinh^{2}\xi(1-\cos(2\eta))\nonumber\\
+B\xi+A_{0}+{\kappa b^{2}\over 4}e^{2(\xi_{0}-\xi)}\cos(2\eta)
\label{psiextsimple}
\end{eqnarray}
In addition, the continuity of $\psi$ on the vortex boundary requires that
$A_{0}$ and $B$ be related by
\begin{equation}
-{1\over 2}{\omega\over 1+q^{2}} a^{2}={1\over 4}\kappa b^{2}+B\xi_{0}+A_{0}
\label{A0B}
\end{equation}
We must also satisfy the continuity of the tangential velocity
${\partial\psi\over\partial\xi}$ at the contact with the vortex. To that
purpose, we need first to express the streamfunction (\ref{psiin}) inside the
vortex in terms of elliptic coordinates. We find:
\begin{equation}
\psi=-{1\over 2}{\omega\over
1+q^{2}}c^{2}(\cosh^{2}\xi\cos^{2}\eta+q^{2}\sinh^{2}\xi\sin^{2}\eta)
\label{psiinell}
\end{equation}
Then, after some algebra, we find that the continuity of the velocity is
satisfied provided that
\begin{equation}
{\kappa\over\omega}={q(1-q)\over 1+q^{2}}
\label{ko}
\end{equation}
and
\begin{equation}
B=-{1\over 2}(\kappa+\omega)q b^{2}
\label{Bannex}
\end{equation}
The relations (\ref{Bannex}) and (\ref{A0B}) just determine the constants
$A_{0}$ and $B$ appearing in the expression (\ref{psiextsimple}) of the
streamfunction. In addition, the condition (\ref{ko}) must be satisfied for a
solution to exist. In a given shear, equation (\ref{ko}) imposes a relation
between the vorticity and the aspect ratio of the vortex.

Regrouping all the results, we find that the streamfunction can be expressed
inside the vortex ($\xi\le\xi_{0}$) by:
\begin{equation}
\psi_{in}={\kappa b^{2}\over 2
q}(q+1)(\cosh^{2}\xi\cos^{2}\eta+q^{2}\sinh^{2}\xi\sin^{2}\eta)
\label{psiinellfin}
\end{equation}
and outside the vortex ($\xi\ge\xi_{0}$) by:
\begin{eqnarray}
\psi_{out}={\kappa\over 4}b^{2}(q^{2}-1)\sinh^{2}\xi(1-\cos(2\eta))\nonumber\\
+{1\over 4}b^{2}\kappa{q+1\over q-1}+{1\over 2}b^{2}\kappa {q+1\over
q-1}(\xi-\xi_{0})\nonumber\\ 
+{\kappa b^{2}\over 4}e^{2(\xi_{0}-\xi)}\cos
(2\eta)
\label{psioutellfin}
\end{eqnarray}

\section{Gravitational instability of a turbulent rotating disk}
\label{sec_turbulentdisk}

In this appendix, we derive an instability criterion for a turbulent rotating
disk in the approximation where the disk is a sheet of zero thickness with
constant mean density $\overline{\sigma}$ and constant angular velocity
$\Omega$. Our study is inspired by the work of Chandrasekhar (1951) concerning
the stability of an infinite homogeneous turbulent medium. To our knowledge, the
results presented in this appendix are new.

The analysis is easiest if we work in a rotating frame of reference. The
equations of the problem are provided by the equation of continuity, the
equation of motion and Poisson equation:
\begin{equation}
{\partial\sigma\over\partial t}+{\partial\over\partial x_{i}}(\sigma u_{i})=0
\label{continuity}
\end{equation}
\begin{eqnarray}
{\partial \over\partial t}(\sigma u_{i})+{\partial\over\partial x_{j}}(\sigma
u_{i}u_{j})=-{\partial p\over\partial x_{i}}-2\Omega \sigma u_{\perp
i}\nonumber\\ 
+\sigma\Omega^{2}x_{i}-\sigma {\partial\Phi\over\partial x_{i}}
\label{pfd}
\end{eqnarray}
\begin{equation}
\Delta\Phi=4\pi G\sigma \delta (z)
\label{newton}
\end{equation}
The symbols  have their usual meaning and  ${\bf u}_{\perp}$ is the vector ${\bf
u}$ rotated by ${\pi\over 2}$. The velocity field $u_{i}=(u_{x},u_{y})$ is
purely two-dimensional  and, in equations (\ref{continuity})(\ref{pfd}), there
is summation over repeated indices. In equation (\ref{newton}), the
$\delta$-function insures that the disk is infinitely thin.   

We assume that the disk is turbulent and write 
\begin{equation}
\sigma=\overline{\sigma}+\delta \sigma,\quad  p=\overline{p}+\delta p,\quad
\Phi=\overline{\Phi}+\delta\Phi
\label{decomposition}
\end{equation}
where $\overline{\sigma}$, $\overline{p}$ and $\overline{\Phi}$ are certain
constants. With the further assumption that the disk is barotropic, i.e
$p=p(\sigma)$, we have
\begin{equation}
{\partial p\over\partial x_{i}}=c_{s}^{2} {\partial\sigma\over\partial x_{i}}
\label{barotropic}
\end{equation}
where $c_{s}=(d p/ d\sigma)^{1/2}$ denotes the velocity of sound. Then, we
invoke Jeans swindle (see, e.g, Binney \& Tremaine, 1987) to eliminate the
centrifugal term, i.e we write
\begin{equation}
{\partial\Phi\over\partial x_{i}}-\Omega^{2}
x_{i}={\partial\delta\Phi\over\partial x_{i}}
\label{swindle}
\end{equation}
This process is permissible if we assume that the centrifugal force is balanced
by a gravitational force that is produced by some unspecified mass distribution
(recall that for an infinite uniform disk, $\nabla\overline{\Phi}$ is
necessarily vertical and cannot, by itself, compensate the centrifugal force).
In our situation, the external force is provided by the sun's gravity. With the
further approximation
\begin{equation}
\sigma {\partial\delta\Phi\over\partial x_{i}}=(\overline{\sigma}+\delta\sigma)
{\partial\delta\Phi\over\partial x_{i}}\simeq \overline{\sigma}
{\partial\delta\Phi\over\partial x_{i}} 
\label{appr}
\end{equation}
the equations of the problem become
\begin{equation}
{\partial\sigma\over\partial t}+{\partial\over\partial x_{i}}(\sigma u_{i})=0
\label{continuityapp}
\end{equation}
\begin{equation}
{\partial \over\partial t}(\sigma u_{i})+{\partial\over\partial x_{j}}(\sigma
u_{i} u_{j})=- c_{s}^{2}{\partial \sigma\over\partial x_{i}}-2\Omega \sigma
u_{\perp i}-\overline{\sigma} {\partial\delta\Phi\over\partial x_{i}}
\label{pfdapp}
\end{equation}
\begin{equation}
\Delta\delta\Phi=4\pi G\delta\sigma \delta (z)
\label{newtonapp}
\end{equation}

From the continuity equation (\ref{continuityapp}), we readily establish that
\begin{equation}
{\partial\over\partial t}\overline{\sigma\sigma'}=-{\partial\over\partial
x_{i}}\overline{\sigma'\sigma  u_{i}}-{\partial\over\partial
x_{i}'}\overline{\sigma\sigma'  u_{i}'}
\label{correq}
\end{equation}
where $\sigma$ and $\sigma'$ are the values of the surface density in ${\bf x}$
and ${\bf x}'$ respectively. We introduce the correlation function of the
density fluctuations, defined by:  
\begin{equation}
C(\xi,t)=\overline{\delta\sigma\delta\sigma'}=
\overline{(\sigma-\overline{\sigma})(\sigma'-\overline{\sigma})}=
\overline{\sigma\sigma'}-\overline{\sigma}^{2}
\label{corr}
\end{equation}
In homogeneous and isotropic turbulence, $C(\xi,t)$ is a scalar function
depending, apart from time, only on the relative distance $\xi=|{\bf x}'-{\bf
x}|$ between the points under consideration. Similarly, we define the quantity
\begin{equation}
L_{i}(\xi,t)=\overline{\sigma'\sigma  u_{i}}=-\overline{\sigma\sigma'  u_{i}'}
\label{L}
\end{equation}
and set
\begin{equation}
{\mb \xi}={\bf x}'-{\bf x}
\label{xiann}
\end{equation}
Remembering that $\partial/\partial x_{i}'=-\partial/\partial
x_{i}=\partial/\partial\xi_{i}$, we can rewrite equation (\ref{correq}) under
the form
\begin{equation}
{\partial C\over\partial t}=2\nabla_{\xi}{\bf L}
\label{correqL}
\end{equation}
An equation of motion for ${\bf L}(\xi,t)$ can be derived in the following
manner. From equations (\ref{continuityapp}) and (\ref{pfdapp}), it is easy to
establish that 
\begin{eqnarray}
{\partial\over\partial t}\overline{\sigma'\sigma u_{i}}+{\partial\over\partial
x_{j}'}\overline{\sigma\sigma' u_{i}u_{j}'}+{\partial\over\partial
x_{j}}\overline{\sigma'\sigma u_{i}u_{j}}\nonumber\\
=-c_{s}^{2}{\partial\overline{\sigma'\sigma}\over\partial
x_{i}}-\overline{\sigma}{\partial\overline{\sigma'\delta\Phi}\over\partial
x_{i}}-2\Omega \overline{\sigma'\sigma u_{\perp i}}
\label{corrLexp}
\end{eqnarray}
or, with more convenient notations
\begin{equation}
{\partial L_{i}\over\partial t}+{\partial\over\partial
\xi_{j}}\overline{\sigma\sigma' (u_{i}u_{j}'- u_{i}u_{j})} =c_{s}^{2}{\partial
C\over\partial \xi_{i}}+\overline{\sigma}{\partial\psi\over\partial
\xi_{i}}-2\Omega L_{\perp i}
\label{corrLnot}
\end{equation}
where we have written
\begin{equation}
\psi=\overline{\sigma'\delta\Phi}=\overline{\delta\sigma'\delta\Phi}
\label{psiann1}
\end{equation}
The correlation function  $\psi$ is  related to $C$ by the Poisson equation
\begin{equation}
\Delta\psi=4\pi G C\delta(z)
\label{poissonpsi}
\end{equation}
Equations (\ref{correqL}) (\ref{corrLnot}) and (\ref{poissonpsi}) are perfectly
general but the system is not closed since the evolution of ${\rm L}(\xi,t)$
involves the fourth-order correlation functions
$\overline{\sigma\sigma'u_{i}u_{j}'}$ and $\overline{\sigma\sigma'u_{i}u_{j}}$
which are not known. To go further, we shall suppose that these fourth-order
correlations can be expressed in terms of the second-order correlations as in a
joint gaussian distribution. For our purposes, this approximation is reasonable
and should provide ample accuracy. Then, by virtue of Wick's theorem, we have:
\begin{eqnarray}
\overline{\sigma\sigma'u_{i}u_{j}}=\overline{\sigma\sigma'}\
\overline{u_{i}u_{j}}+\overline{\sigma u_{i}}\ \overline{\sigma'
u_{j}}+\overline{\sigma u_{j}}\ \overline{\sigma' u_{i}}\nonumber\\
=\overline{\sigma\sigma'}{1\over 2}\overline{u^{2}}\delta_{ij}
\label{wick1}
\end{eqnarray}
\begin{eqnarray}
\overline{\sigma\sigma'u_{i}u_{j}'}=\overline{\sigma\sigma'}\
\overline{u_{i}u_{j}'}+\overline{\sigma u_{i}}\ \overline{\sigma'
u_{j}'}+\overline{\sigma u_{j}'}\ \overline{\sigma' u_{i}}\nonumber\\
=\overline{\sigma\sigma'}\ \overline{u_{i}u_{j}'}-\overline{\sigma u_{i}'}\
\overline{\sigma u_{j}'}
\label{wick2}
\end{eqnarray}
With the foregoing substitution, equation (\ref{corrLnot}) becomes
\begin{eqnarray}
{\partial L_{i}\over\partial t}+{\partial\over\partial
\xi_{j}}\lbrack\overline{\sigma\sigma'}\ \overline{u_{i}u_{j}'}-\overline{\sigma
u_{i}'}\ \overline{\sigma u_{j}'}\rbrack    =(c_{s}^{2}+{\overline{u^{2}}\over
2}){\partial C\over\partial \xi_{i}}\nonumber\\
+\overline{\sigma}{\partial\psi\over\partial \xi_{i}}-2\Omega L_{\perp i}
\label{corrLgauss}
\end{eqnarray}
For sufficiently large values of $\xi$, the terms in the velocity correlations
in equation (\ref{corrLgauss}) will become negligible and equation
(\ref{corrLgauss}) will tend to
\begin{equation}
{\partial {\bf L}\over\partial t} =(c_{s}^{2}+{\overline{u^{2}}\over
2})\nabla_{\xi} C+\overline{\sigma}\nabla_{\xi}\psi-2{\mb \Omega}\wedge {\bf L}
\label{corrLgaussinf}
\end{equation}
Equations (\ref{correqL})(\ref{corrLgaussinf}) and (\ref{poissonpsi}) are
mathematically similar to the linearized equations which appear in the usual
problem of Jeans instability for a thin rotating disk (see, e.g, Binney \&
Tremaine 1987). However, their physical meaning is different since equations
(\ref{correqL})(\ref{corrLgaussinf}) and (\ref{poissonpsi}) have been derived
explicitely from the analysis of the correlations in a turbulent medium. 

These equations admit sound waves of the form 
\begin{equation}
C=\hat C e^{-i\omega t}J_{0}(k\xi)
\label{Cann}
\end{equation}
\begin{equation}
{\bf L}=\hat {\bf L} e^{-i\omega t}J_{0}(k\xi)
\label{Lann}
\end{equation}
\begin{equation}
\psi=\hat \psi e^{-i\omega t}J_{0}(k\xi)e^{-k|z|}
\label{psiann}
\end{equation}
where $J_{0}$ is Bessel function of order zero. The solution (\ref{psiann})
satisfies the Laplace equation $\Delta\psi=0$ for $z\neq 0$ [see equation
(\ref{poissonpsi})]. In the plane $z=0$, we have $\psi=\hat \psi e^{-i\omega
t}J_{0}(k\xi)$. Equation (\ref{poissonpsi}) implies that $\hat \psi$ must be
related to $\hat C$ in a special manner. To see that, we integrate the Poisson
equation  from $-\zeta$ to $+\zeta$ (where $\zeta$ is a positive constant) and
let $\zeta\rightarrow 0$. Since ${\partial^{2}\psi\over\partial x^{2}}$ and
${\partial^{2}\psi\over\partial y^{2}}$ are continuous at $z=0$, but
${\partial^{2}\psi\over\partial z^{2}}$ is not, we have:
\begin{eqnarray}
\lim_{\zeta\rightarrow 0}\int_{-\zeta}^{\zeta}{\partial^{2}\psi\over\partial
z^{2}}dz=\lim_{\zeta\rightarrow 0}\biggl \lbrack {\partial\psi\over\partial
z}\biggr \rbrack_{-\zeta}^{\zeta}\nonumber\\ 
=4\pi
GC\int_{-\zeta}^{\zeta}\delta(z)dz=4\pi GC
\label{bt}
\end{eqnarray} 
Hence $-2 k \psi=4\pi G C$, or
\begin{equation}
\hat \psi =-{2\pi G\hat C\over k}
\label{psihat}
\end{equation}  
Substituting for $C$, ${\bf L}$ and $\psi$ in equations
(\ref{correqL})(\ref{corrLgaussinf}) and (\ref{poissonpsi}), we obtain
\begin{equation}
-i\omega C=2{1\over\xi}{\partial\over\partial\xi}(\xi L_{\xi})
\label{Comega}
\end{equation} 
\begin{equation}
-i\omega L_{\xi}=(c_{s}^{2}+{\overline{u^{2}}\over 2}){\partial C\over\partial
\xi}+\overline{\sigma} {\partial\psi\over\partial\xi}+2\Omega L_{\eta}
\label{Lxi}
\end{equation}
\begin{equation}
-i\omega L_{\eta}=-2\Omega L_{\xi}
\label{Leta}
\end{equation}
where $\xi,\eta$ are polar coordinates. Eliminating $L_{\eta}$ between
(\ref{Lxi}) and (\ref{Leta}) yields
\begin{equation}
{\omega^{2}-4\Omega^{2}\over i\omega}L_{\xi}=(c_{s}^{2}+{\overline{u^{2}}\over
2}){\partial C\over\partial \xi}+\overline{\sigma}
{\partial\psi\over\partial\xi}
\label{Lxiexpli}
\end{equation} 
Substituting the foregoing expression for $L_{\xi}$ into equation
(\ref{Comega}), we arrive at
\begin{equation}
(4\Omega^{2}-\omega^{2})C=2(c_{s}^{2}+{\overline{u^{2}}\over 2})\Delta_{\xi}C +2
\overline{\sigma} \Delta_{\xi}\psi
\label{dispdiff}
\end{equation} 
Using the expressions (\ref{Cann}) (\ref{psiann}) (\ref{psihat}) for $C$ and
$\psi$ and remembering that $J_{0}(k\xi)$ is an eigenfunction of the
two-dimensional Laplacian operator with eigenvalue $-k^{2}$, we obtain the
dispersion relation
\begin{equation}
\omega^{2}=4\Omega^{2}+2\biggl \lbrack k^{2}(c_{s}^{2}+{\overline{u^{2}}\over
2})-2\pi G \overline{\sigma} k\biggr\rbrack
\label{dispann}
\end{equation} 
This result differs from the usual dispersion relation (\ref{dispersion}) in the
occurence of $c_{s}^{2}+{\overline{u^{2}}\over 2}$ in place of $c_{s}^{2}$ and
by a factor $2$. Similar differences with the usual Jeans dispersion relation
were noticed by Chandrasekhar (1951) in his analysis of an infinite homogeneous
turbulent medium.  

The function $\omega^{2}(k)$ is quadratic in $k$ with a minimum at $k_{c}=\pi
G\overline{\sigma}/( c_{s}^{2}+{\overline{u^{2}}\over 2})$. The disk will be
unstable to some wavelengh $k^{-1}$ provided that $\omega^{2}(k_{c})<0$, i.e:
\begin{equation}
\sqrt{2\biggl (c_{s}^{2}+{\overline{u^{2}}\over 2}\biggr )}<{\pi
G\overline{\sigma}\over \Omega}
\label{instab}
\end{equation}
This is the generalisation of the Toomre instability criterion (\ref{jeans1}) in
the case where the disk is turbulent. For a real disk, with finite thickness
$H$, ${\overline{u^{2}}\over 2}$ should be replaced by ${\overline{u^{2}}\over
3}$ since small-scale fluctuations are basically three-dimensional. When the
turbulent dispersion $c_{turb}=\sqrt{\overline{u^{2}}}$ dominates over the sound
velocity, as it is the case in our study, we obtain
\begin{equation}
c_{turb}<{\pi G\overline{\sigma}\over \Omega}
\label{instabapp}
\end{equation}  
This result justifies the procedure used in section \ref{sec_application2} of
replacing the sound velocity occuring in equation (\ref{jeans1}) by the more
relevant turbulent dispersion of the particles. However, in more general
situations, the turbulent dispersion needs not be large compared to the velocity
of sound and the criterion (\ref{instab}) should be used.

--------------------------------------------

\end{document}